\documentclass{article}
\usepackage[final]{neurips_2021}
\usepackage[utf8]{inputenc} 
\usepackage[T1]{fontenc}    
\usepackage[]{hyperref}       
\usepackage{booktabs}       
\usepackage{amsfonts}       
\usepackage{nicefrac}       
\usepackage{microtype}      
\usepackage{nicefrac}
\usepackage{amssymb}
\usepackage{amsmath}
\usepackage{amsthm}
\usepackage{MnSymbol} 
\newcommand{\grad}{\nabla}
\usepackage{url}
\usepackage{lettrine}
\usepackage{framed}
\usepackage{color}
\usepackage{algorithm}
\usepackage{algorithmic}
\usepackage{overpic}
\usepackage{dsfont}
\usepackage{booktabs} 
\usepackage{bm}
\usepackage{bbm}
\usepackage{ifthen}
\usepackage{graphicx}
\usepackage{dsfont}
\usepackage{framed}
\usepackage{overpic}
\usepackage[svgnames]{xcolor}
\newtheorem{theorem}{Theorem}
\newtheorem{proposition}{Proposition}
\newtheorem{definition}{Definition}
\newtheorem{lemma}{Lemma}
\usepackage{xr}
\usepackage{subfigure}
\graphicspath{{../ICML2021/codes/plots/}}
\hypersetup{
	colorlinks=true,
	linkcolor=blue,
	urlcolor=red,
	citecolor=DarkBlue,
	linkbordercolor={0 0 1}
}

\newcommand{\alglinelabel}{%
  \addtocounter{ALC@line}{-1}
  \refstepcounter{ALC@line}
  \label
}

\renewcommand{\algorithmicrequire}{\textbf{Input:}}
\renewcommand{\algorithmicensure}{\textbf{Output:}}

\title{\texttt{LeadCache}: Regret-Optimal Caching in Networks}
\author{
  Debjit Paria\thanks{Work done at the Indian Institute of Technology Madras as a part of the first author's Master's thesis.} \\
  Department of Computer Science\\
  Chennai Mathematical Institute\\
  Chennai 603103, India\\
  \texttt{debjit.paria1999@gmail.com} \\
   \And
   Abhishek Sinha \\
   Department of Electrical Engineering \\
  Indian Institute of Technology Madras \\
  Chennai 600036, India\\
   \texttt{abhishek.sinha@ee.iitm.ac.in} 
}

\begin{document}

\maketitle

\begin{abstract} 
We consider an online prediction problem in the context of network caching.
Assume that multiple users are connected to several caches via a bipartite network. At any time slot, each user may request an arbitrary file chosen from a large catalog. 
A user's request at a slot is met if the requested file is cached in at least one of the caches connected to the user. Our objective is to predict, prefetch, and optimally distribute the files on the caches at each slot to maximize the total number of cache hits. The problem is non-trivial due to the non-convex and non-smooth nature of the objective function. In this paper, we propose \texttt{LeadCache} - an efficient online caching policy based on the Follow-the-Perturbed-Leader paradigm. We show that \texttt{LeadCache} is regret-optimal up to a factor of $\tilde{O}(n^{3/8}),$ where $n$ is the number of users. We design two efficient implementations of the \texttt{LeadCache} policy, one based on Pipage rounding and the other based on Madow's sampling, each of which makes precisely one call to an LP-solver per iteration.
Furthermore, with a Strong-Law-type assumption, we show that the total number of file fetches under \texttt{LeadCache} remains almost surely finite over an infinite horizon. Finally, we derive an approximately tight regret lower bound using results from graph coloring. We conclude that the learning-based \texttt{LeadCache} policy decisively outperforms the state-of-the-art caching policies both theoretically and empirically.

\end{abstract}


\section{Introduction}
\begin{figure}
\centering
	\begin{minipage}[b]{0.45\linewidth}
	\begin{overpic}[width=0.85\textwidth]{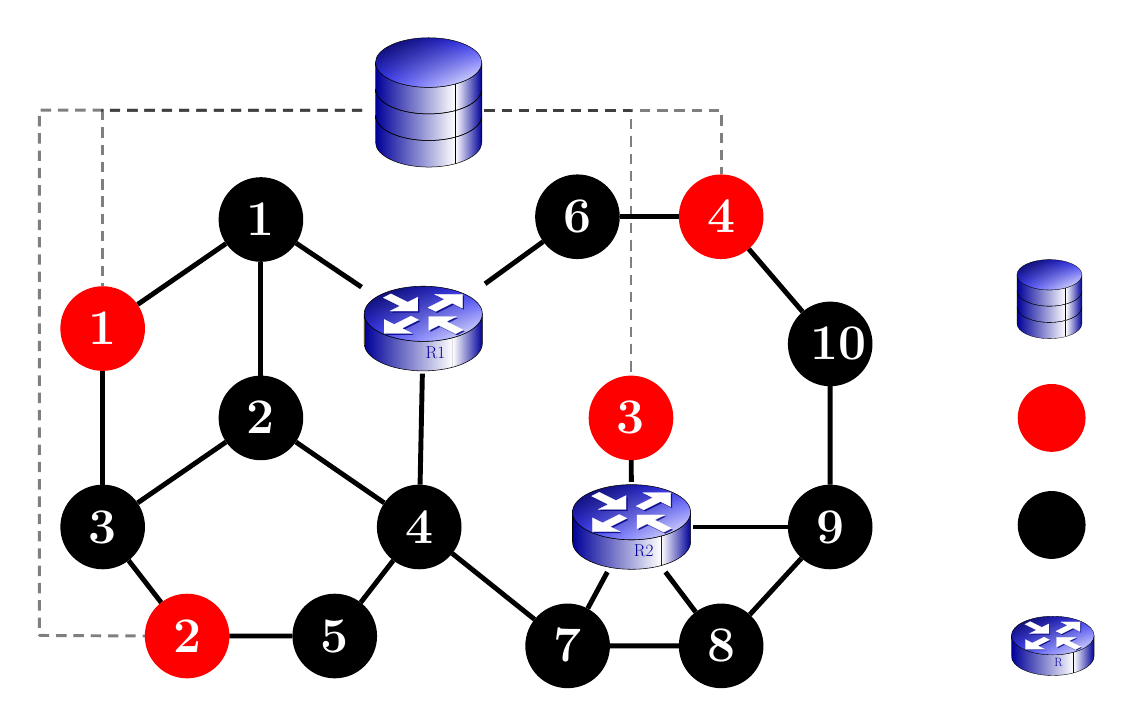}
	\put(100,26){\footnotesize{Cache}}	
	\put(35, -2){\small{(a)}}
	\put(100, 41){\footnotesize{Remote}}
	\put(100,35){\footnotesize{server}}
	\put(100,16){\footnotesize{User}}
	\put(100,6){\footnotesize{Router}}
	\end{overpic}
\end{minipage}
\hspace{10pt}
	\begin{minipage}[b]{0.45\linewidth}
	\centering
	\begin{overpic}[width=0.45\textwidth]{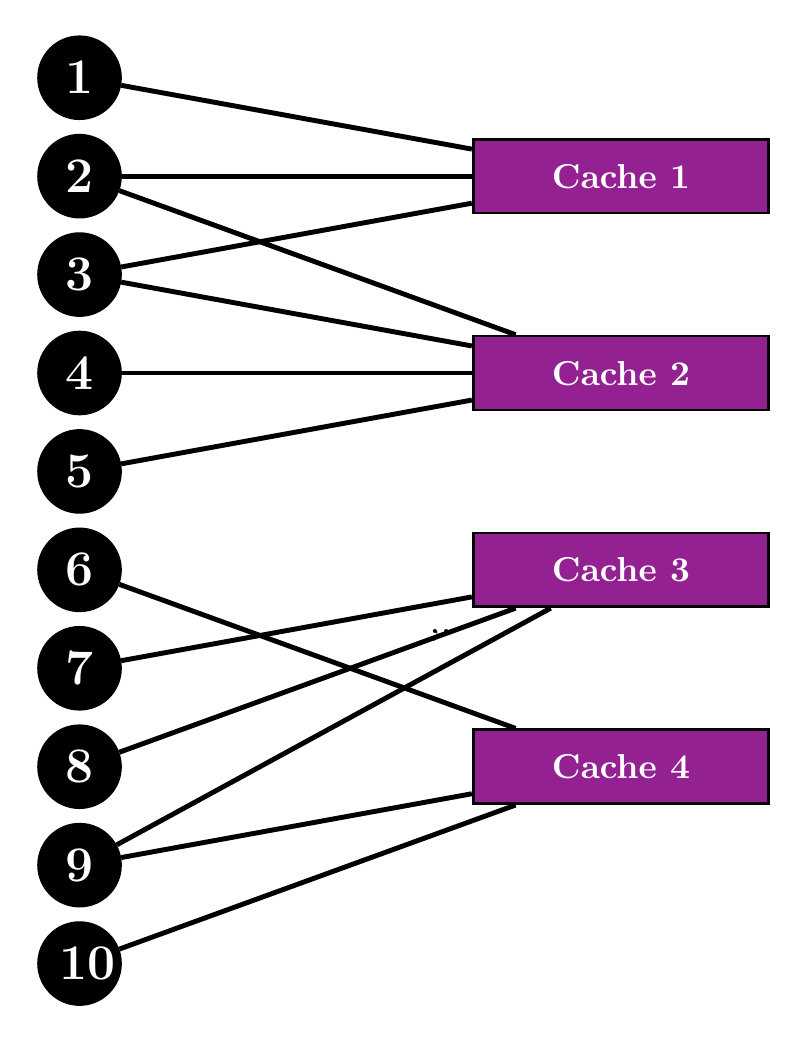}
	\put(-32,50){\footnotesize{$n$ users}}	
	\put(-40,42){\footnotesize{max-degree}}
	\put(-25, 32){\footnotesize{$=\Delta$}}
	\put(40, -2){\small{(b)}}
	\put(45,10){\footnotesize{$m$ \textrm{caches}}}
     \put(32,95){\footnotesize{max-degree = $d$}}
	\end{overpic}
	\end{minipage}
	\caption{Reduction of the \textsf{Network Caching} problem (a) to the \textsf{Bipartite Caching} problem (b). In this schematic, we assumed that a cache, located within two hops, is reachable to a user.}
	\label{reduction_fig}
\end{figure}
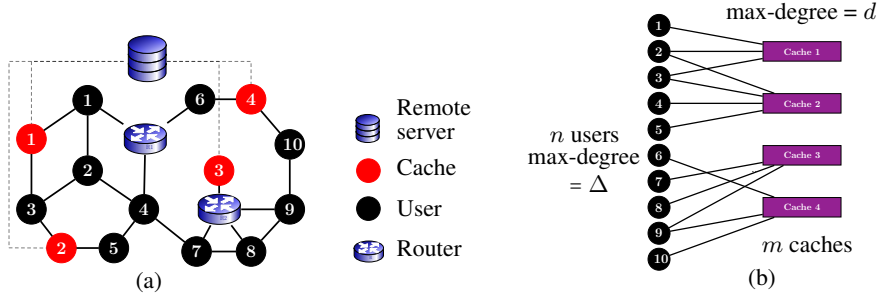

We consider an online structured learning problem, called \textsf{Bipartite Caching}, that lies at the core of many large-scale internet services, including Content Distribution Networks (CDN) and Cloud Computing. Formally, a set $\mathcal{I}$ of $n$ users is connected to a set $\mathcal{J}$ of $m$ caches via a bipartite network $G(\mathcal{I} \cupdot \mathcal{J}, E)$. Each cache is connected to at most $d$ users, and each user is connected to at most $\Delta$ caches (see Figure \ref{reduction_fig} (b)). There is a catalog consisting of $N$ unique files, and each of the $m$ caches can host at most $C$ files at a time (in practice, $C \ll N$). 
The system evolves in discrete time slots. Each of the $n$ users may request any file from the catalog at each time slot. The file requests could be dictated by an adversary. Given the storage capacity constraints, an online caching policy decides the files to be cached on different caches at each slot before the requests for that slot arrive.   
The objective is to maximize the total number of \emph{hits} by the unknown incoming requests by coordinating the caching decisions among multiple caches in an online fashion. The \textsf{Bipartite Caching} problem is a strict generalization of the 
online $\bm{k}$-\textbf{sets} problem that predicts a set of $k$ items at each round so that the predicted set includes the item chosen by the adversary \citep{koolen2010hedging, cohen2015following}. However, unlike the $\bm{k}$-\textbf{sets} problem, which predicts a single subset at a time, in this problem, we are interested in sequentially predicting multiple subsets, each corresponding to one of the caches. The interaction among the caches through the non-linear reward function makes this problem challenging.  

The \textsf{Bipartite Caching} problem is a simplified abstraction of the more general \textsf{Network Caching} problem central to the commercial CDNs, such as Akamai \hbox{\citep{akamai}}, Amazon Web Services (AWS), and Microsoft Azure \citep{CIT-104}. In the \textsf{Network Caching} problem, one is given an arbitrary graph $\mathcal{G}(V,E)$, a set of users $\mathcal{I}\subseteq V$, and a set of caches $\mathcal{J}\subseteq V$. A user can retrieve a file from a cache only if the cache hosts the requested file. If the $i$\textsuperscript{th} user retrieves the requested file from the $j$\textsuperscript{th} cache, the user receives a reward of $r_{ij} \geq 0$ for that slot. If the requested file is not hosted in any of the caches reachable to the user, the user receives zero rewards for that slot. 
  The goal of a network caching policy is to dynamically place files on the caches so that cumulative reward obtained by all users is maximized. The \textsf{Network Caching} problem reduces to the \textsf{Bipartite Caching} problem when the rewards are restricted to the set $\{0,1\}$. It will be clear from the sequel that the algorithms presented in this paper can be extended to the general \textsf{Network Caching} problem as well. 


\subsection{Problem Formulation} \label{action}
Denote the file requested by the $i$\textsuperscript{th} user by the one-hot encoded $N$-dimensional vector $\bm{x}_t^i.$ In other words, $\bm{x}^i_{tf}=1$ if the $i$\textsuperscript{th} user requests file $f \in [N]$ at time slot $t$, or $\bm{x}^i_{tf}=0$ otherwise. Since a user may request at most one file per time slot, we have:
$\sum_{f=1}^N \bm{x}^i_{tf} \leq 1, ~~\forall i \in \mathcal{I}, \forall t. $
An online caching policy prefetches files on the caches at every time slot based on past requests. Unlike classical caching policies, such as \textsf{LRU, LFU, FIFO, Marker}, that fetch a file immediately upon a cache-miss, we do not enforce this constraint in the problem statement.   
 The set of files placed on the $j$\textsuperscript{th} cache at time $t$ is represented by the $N$-dimensional incidence vector $\bm{y}_t^j \in \{0,1\}^N.$ In other words, $\bm{y}_{tf}^j = 1$ if the $j$\textsuperscript{th} cache hosts file $f \in [N]$ at time $t$, or $\bm{y}_{tf}^j = 0$ otherwise. Due to cache capacity constraints, the following inequality must be satisfied at each time slot $t$:
 $\sum_{f =1}^N \bm{y}^j_{tf} \leq C, ~~ \forall j \in \mathcal{J}.$
 
The set of all admissible caching configurations, denoted by $\mathcal{Y} \subseteq \{0,1\}^{Nm},$ is dictated by the cache capacity constraints. In principle, the caching policy is allowed to replace all elements of the caches at every slot, incurring a potentially huge downloading cost over an interval. However, in Section \ref{fetch}, we show that the total number of files fetched to the caches under the proposed \texttt{LeadCache} policy remains almost surely finite under very mild assumptions on the file request process.  

The $i$\textsuperscript{th} user receives a \emph{cache hit} at time slot $t$ if and only if \emph{any} of the caches connected to the $i$\textsuperscript{th} user hosts the file requested by the user at slot $t$. In the case of a cache hit, the user obtains a unit reward. On the other hand, in the case of a \emph{cache miss}, the user receives zero rewards for that slot. Hence, for a given aggregate request vector from all users $\bm{x}_t= \big( \bm{x}_t^i, i \in \mathcal{I}\big) $ and the aggregate cache configuration vector of all caches $\bm{y}_t =\big( \bm{y}_t^j, j \in \mathcal{J}\big)$, the total reward $q(\bm{x}_t, \bm{y}_t)$ obtained by the users at time $t$ may be expressed as follows:
\begin{eqnarray} \label{reward_definition}
	q(\bm{x}_t, \bm{y}_t) \equiv  \sum_{i \in \mathcal{I}} {\bm{x}_t^i} \cdot \min \bigg\{\bm {1}_{N\times 1},\big(\sum_{j \in \partial^+(i)} \bm{y}_t^j\big)\bigg\},
\end{eqnarray}
where $\bm{a} \cdot \bm{b}$ denotes the inner-product of the vectors $\bm{a}$ and $\bm{b},$ $\bm{1}_{N \times 1}$ denotes the $N$-dimensional all-one column vector, the set $\partial^+(i)$ denotes the set of all caches connected to the $i$\textsuperscript{th} user, and the ``$\min$" operator is applied component wise. 
The total reward $Q(T)$ accrued in a time-horizon of length $T$ is obtained by summing the slot-wise rewards, \emph{i.e.,} $Q(T)=\sum_{t=1}^Tq(\bm{x}_t, \bm{y}_t).$ Following the standard practice in the online learning literature, we measure the performance of any online policy $\pi$ using the notion of (static) \emph{regret} $R^\pi(T)$, defined as the maximum difference in the cumulative rewards obtained by the optimal fixed caching-configuration in hindsight and that of the online policy $\pi$, \emph{i.e.,}
\begin{eqnarray} \label{regret_def}
	R^\pi(T) \stackrel{(\textrm{def.})}{=}\sup_{\{\bm{x}_t\}_{t=1}^T}\bigg(\sum_{t=1}^T q(\bm{x}_t, \bm{y}^*) - \sum_{t=1}^T q(\bm{x}_t, \bm{y}_t^\pi) \bigg),
\end{eqnarray}
 where $\bm{y}^*$ is the best static cache-configuration in \emph{hindsight} for the file request sequence $\{\bm{x}_t\}_{t=1}^T$, \emph{i.e.,} $\bm{y}^* = \arg\max_{\bm{y} \in\mathcal{Y}} \sum_{t=1}^T q(\bm{x}_t, \bm{y})$. We assume that the file request sequence is generated by an \emph{oblivious adversary}, \emph{i.e.,}
the entire request sequence $\{\bm{x}_t\}_{t \geq 1}$ is fixed a priori. 
 Note that the problem is \emph{non-convex}, as we seek binary cache allocations.
With an eye towards efficient implementation, later we will also consider the problem of  designing efficient policies that guarantee a sub-linear $\alpha$-regret for a suitable value of $\alpha <1$ 
\citep{garber2021efficient, kakade2009playing, fujita2013combinatorial}.
\section{Background and Related Work} \label{lit}
Online Linear Optimization (OLO) is a canonical online learning problem that can be formulated as a repeated game played between a learner (also known as the forecaster) and an adversary \citep{cesa2006prediction}. In this model, at every time slot $t$, the policy selects an action $\bm{y}_t$ from a  feasible set $\mathcal{Y} \subseteq \mathbb{R}^d.$ After that, the adversary reveals a reward vector $\bm{x}_t$ from a set $\mathcal{X} \subseteq \mathbb{R}^d.$ The adversary is assumed to be \emph{oblivious}, \emph{i.e.,} the sequence of reward vectors is fixed before the game begins. With the above choices, the policy receives a scalar reward $q(\bm{x}_t, \bm{y}_t):=\langle \bm{x}_t, \bm{y}_t \rangle$ at slot $t$. A classic objective in this setting is to design a policy with a small regret. 
Follow the Perturbed Leader \textsf{(FTPL)}, is a well-known online policy for the OLO problem \citep{hannan1957approximation}. At time slot $t$, the \textsf{FTPL} policy adds a random noise vector $\bm{\gamma}_t$ to the cumulative reward vector $\bm{X}_t = \sum_{\tau=1}^{t-1} \bm{x}_\tau$, and then selects the best action against this perturbed reward, \emph{i.e.,} $\bm{y}_t := \arg\max_{\bm{y} \in \mathcal{Y}}\langle \bm{X}_t+ \bm{\gamma}_t, \bm{y} \rangle.$ 
See \citet{abernethy2016perturbation} for a unifying treatment of the \textsf{FTPL} policies through the lens of stochastic smoothing. 


 
A large number of papers on caching assume some stochastic model for the file request sequence, \emph{e.g.,} Independent Reference Model (IRM) and Shot Noise Model (SNM) \citep{traverso2013temporal}. 
Classic page replacement algorithms, such as \textsf{MIN, LRU, LFU, and FIFO}, are designed to minimize the \emph{competitive ratio} with adversarial requests \citep{borodin2005online, van2007short, lee1999existence, dan1990approximate}. These algorithms, being \emph{non-prefetching} in nature, replace a page on demand upon a cache-miss. However, since the competitive ratio metric is multiplicative in nature, there can be a large gap between the hit ratio of a competitively optimal policy and the optimal offline policy. To design better algorithms, the caching problem has recently been investigated through the lens of regret minimization with \emph{prefetching} policies that \emph{learn} from the past request sequence \citep{vitter1996optimal, krishnan1998optimal}. 
 \citet{daniely2019competitive} considered the problem of minimizing the regret plus the switching cost for a single cache. The authors proposed a variant of the celebrated exponential weight algorithm \citep{littlestone1994weighted, freund1997decision} that ensures the  minimum competitive ratio and a small but sub-optimal regret. 
 The \textsf{Bipartite Caching} model was first proposed in a pioneering paper by \citet{shanmugam2013femtocaching}, where they considered a stochastic version of the problem with known file popularities. \citet{paschos2019learning} proposed an Online Gradient Ascent (OGA)-based \textsf{Bipartite Caching} policy that allows caching a fraction of the Maximum Distance Separable (MDS)-coded files.  Closely related to this paper is the recent work by \citet{SIGMETRICS20}, where the authors designed a regret-optimal single-cache policy and a \textsf{Bipartite Caching} policy for fountain-coded files. However, the fundamental problem of designing a regret-optimal uncoded caching policy for the \textsf{Bipartite Caching} problem was left as an open problem.
 \paragraph{Why standard approaches fail:} A straightforward way to formulate the \textsf{Bipartite Caching} problem is to pose it as an instance of the classic \textsf{Prediction with Expert Advice} problem \citep{cesa2006prediction}, where each of the possible $\binom{N}{C}^m$ cache configurations is treated as experts. However, this approach is computationally infeasible due to the massive number of resulting experts. 
 Furthermore, the expected \textsf{FTPL} policy for convex losses in \citet{hazan2019introduction} and its sampled version in \citet{hazan2020faster} are both computationally intensive.
 In view of the above challenges, we now present our main technical contributions.

\section{Main Results}\label{results} 
\begin{figure}
\centering
	\includegraphics[width=0.5\textwidth]{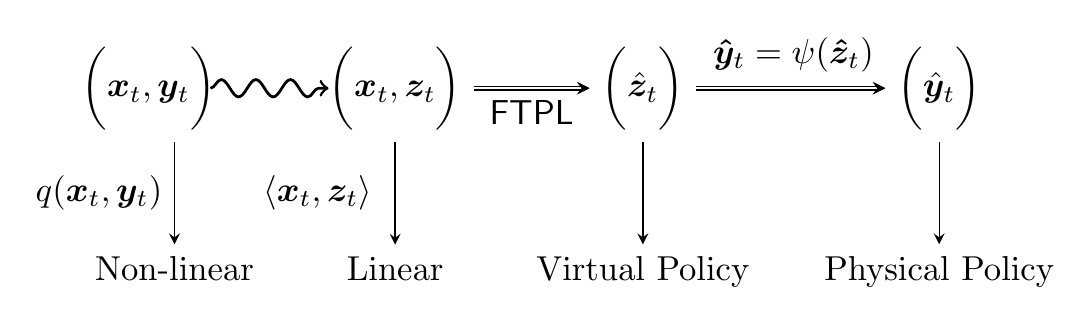}
	\caption{ Reduction pipeline illustrating the translation of the \textsf{Bipartite Caching} problem with a non-linear reward function to an online learning problem with a linear reward function. 
	}
	\label{reduction_figure}
\end{figure}

\paragraph{1. \textsf{FTPL} for a non-linear non-convex problem:}
We propose \texttt{LeadCache}, a network caching policy based on the \textsf{Follow the Perturbed Leader} paradigm. The \emph{non-linearity} of the reward function and the non-convexity of the feasible set (due to the integrality of cache allocations) pose a significant challenge in using the generic \textsf{FTPL} framework \citep{abernethy2016perturbation}. To circumvent this difficulty, 
we switch to a \emph{virtual action} domain $\mathcal{Z}$ where the reward function is linear. 
We use an anytime version of the \textsf{FTPL} policy for designing a virtual policy for the linearized virtual learning problem. 
Finally, we translate the virtual policy back to the original action domain $\mathcal{Y}$ with the help of a mapping $\psi: \mathcal{Z} \to \mathcal{Y},$ obtained by solving a combinatorial optimization problem (see Figure \ref{reduction_figure} for the overall reduction pipeline). 
\paragraph{2. New Rounding Techniques and $\alpha$-regret:}
 The mapping $\psi$, which translates the virtual actions to physical caching actions in the above scheme, turns out to be an \textsf{NP-hard} Integer Linear Program. As our second contribution, we design a linear-time Pipage rounding technique for this problem \citep{ageev2004pipage}. Incidentally, our rounding process substantially improves upon a previous rounding scheme proposed by \citet{shanmugam2013femtocaching} in the context of caching with i.i.d.\ requests.  
  Next, we propose a linear-time randomized rounding scheme that yields an efficient online policy with a provable sub-linear $\alpha \equiv 1-\nicefrac{1}{e}$ regret. The proposed randomized rounding algorithm exploits a classical sampling technique used in statistical surveys.
\paragraph{3. Bounding the Switching Cost:}  As our third contribution, we show that if the file requests are generated by a stochastic process satisfying a mild Strong Law-type property, then the caching configuration under the \texttt{LeadCache} policy converges to the corresponding optimal configuration \emph{almost surely} \emph{in finite time}. As new file fetches to the caches from the remote server consume bandwidth, this result implies that the proposed policy offers the best of both worlds - (1) a sub-linear regret for adversarial requests, and (2) finite downloads for ``stochastically regular" requests.  
\paragraph{4. New Regret Lower Bound:} As our final contribution, we derive minimax regret lower bound that is tight up to a factor of $\tilde{O}(n^{\nicefrac{3}{8}})$. Our lower bound sharpens a result in \cite{SIGMETRICS20}.  
The proof of the lower bound critically utilizes graph coloring theory and the probabilistic Balls-into-Bins framework. 


\section{The \texttt{LeadCache} Policy} \label{achievability}
In this section, we propose \texttt{LeadCache} - an efficient network caching policy that guarantees near-optimal regret.
Since the reward function \eqref{reward_definition} is non-linear, 
 we \emph{linearize} the problem by switching to a \emph{virtual} action domain $\mathcal{Z},$ as detailed below.  

\paragraph{The Virtual Caching Problem:} First, we consider an associated Online Linear Optimization (OLO) problem, called \textsf{Virtual Caching}, as defined next. In this problem, at each slot $t$, a virtual action $\bm{z}_t \equiv \big( \bm{z}_t^i, i \in \mathcal{I}\big)$ is taken in response to the file requests received so far. The $i$\textsuperscript{th} component of the virtual action, denoted by $\bm{z}_t^i \in \{0,1\}^N,$ 
roughly indicates the availability of the files in the caches connected to the $i$\textsuperscript{th} user. The set of all admissible virtual actions, denoted by $\mathcal{Z} \subseteq \{0,1\}^{N\times n}$, is defined below in Eqn.\ \eqref{virtual_physical}. The reward $r(\bm{x}_t, \bm{z}_t)$ accrued by the virtual action $\bm{z}_t$ for the file request vector $\bm{x}_t$ at the $t$\textsuperscript{th} slot is given by their inner product, \emph{i.e.,}
\begin{eqnarray} \label{reward_definition2}
  		r(\bm{x}_t, \bm{z}_t) := \langle \bm{x}_t, \bm{z}_t \rangle = \sum_{i \in \mathcal{I}} \bm{x}_t^i \cdot \bm{z}_t^i.
  \end{eqnarray}

\textbf{Virtual Actions:} The set $\mathcal{Z}$ of all admissible virtual actions is defined as the set of all binary vectors $\bm{z} \in \{0,1\}^{N\times n}$ such that the following  component wise inequalities hold for some admissible physical cache configuration vector $\bm{y} \in \mathcal{Y}:$ 
\begin{eqnarray} \label{virtual_physical}
\bm{z}^i\leq \min\big\{ \bm{1}_{N\times 1}, \big(\sum_{j \in \partial^+(i)}\bm{y}^j\big) \big\}, ~~1 \leq i \leq n.
\end{eqnarray}
More explicitly, the set $\mathcal{Z}$ can be characterized as the set of all binary vectors $\bm{z} \in \{0,1\}^{N\times n}$ satisfying the following constraints for some feasible $\bm{y} \in \mathcal{Y}$:
\begin{eqnarray}  \label{virtual_Z}
z^i_f &\leq& \sum_{j \in \partial^+(i)} y^j_f, ~~\forall i \in \mathcal{I},f \in [N]	\label{constr1}\\
\sum_{f=1}^N y_f^j & \leq & C, ~~ \forall j \in \mathcal{J},	\label{cap_ctr2}\\
 y^j_f, z^i_f &\in& \{0,1\}, ~~\forall i\in \mathcal{I}, \forall j \in \mathcal{J}, f \in [N] \label{bin_constr1}. 
\end{eqnarray}
%
 Let $\psi : \mathcal{Z} \to \mathcal{Y}$ be a mapping that maps any admissible virtual action $\bm{z} \in \mathcal{Z}$ to a corresponding physical caching action $\bm{y} $ satisfying the condition  \eqref{virtual_physical}. Hence, the binary variable $z^i_{tf}=1$ \emph{only if} the file $f$ is hosted in one of the caches connected to the $i$\textsuperscript{th} user at time $t$ in the physical configuration $\bm{y}_t=\psi(\bm{z}_t)$. The mapping $\psi$ may be used to translate any virtual caching policy $\pi^{\textrm{virtual}}=\{\bm{z}_t\}_{t \geq 1},$ to a physical caching policy $\pi^{\textrm{phy}}\equiv \psi(\pi^{\textrm{virtual}})=\{\bm{y}_t\}_{t \geq 1}$ through the correspondence $\bm{y}_t = \psi (\bm{z}_t), \forall t\geq 1.$
The following lemma relates the regrets incurred by these two online policies:
  \begin{lemma} \label{regret_comp}
  For any virtual caching policy $\pi^{\textrm{virtual}}$, define a physical caching policy $\pi^{\textrm{phy}} = \psi(\pi^{\textrm{virtual}})$ as above. Then the regret of the policy $\pi^{\textrm{phy}}$ is bounded above by that of the policy $\pi^{\textrm{virtual}}$, \emph{i.e.,}
 $$ R^{\pi^{\textrm{phy}}}_T \leq  R^{\pi^{\textrm{virtual}}}_T, ~~~\forall T\geq 1. 	$$
	 \end{lemma}
Please refer to Section \ref{regret_comp_proof} in the supplementary material for the proof of Lemma \ref{regret_comp}.
Lemma \ref{regret_comp} implies that any low-regret virtual caching policy may be used to design a low-regret physical caching policy using the non-linear mapping $\psi(\cdot)$. The key advantage of the virtual caching problem is that it is a standard OLO problem. Hence, in our proposed \texttt{LeadCache} policy, we use an anytime version of the \textsf{FTPL} policy for solving virtual caching problem. The overall \texttt{LeadCache} policy is described below in Algorithm \ref{uncoded_inelastic}:
\begin{algorithm}
\caption{The \texttt{LeadCache} Policy}
\label{uncoded_inelastic}
\begin{algorithmic}[1]
\STATE $\bm{X}(0)\gets \bm{0}$
\STATE Sample $\bm{\gamma} \stackrel{\textrm{i.i.d.}}{\sim} \mathcal{N}(0,\mathsf{1}_{Nn\times 1})$ 
\FOR {$t=1$ to $T$}
\STATE $\bm{X}(t) \gets \bm{X}(t-1) + \bm{x}_t$
\STATE $\eta_t \gets\frac{n^{3/4}}{(2d(\log \frac{N}{C} + 1))^{1/4}}\sqrt{\frac{t}{Cm}}$
\STATE $\bm{\Theta}(t)\gets \bm{X}(t)+ \eta_t\bm{\gamma}$ \alglinelabel{addG}
\STATE $\bm{z}_t \gets \max_{\bm{z} \in \mathcal{Z}} \langle \bm{\Theta}(t), \bm{z}\rangle.$ \alglinelabel{algo_step}	
\STATE $\bm{y}_t \gets \psi(\bm{z}_t).$\alglinelabel{algo_step2}
  \ENDFOR
\end{algorithmic}
\end{algorithm}

In Algorithm \ref{uncoded_inelastic}, the flattened $Nn \times 1$ dimensional vector $\bm{X}(t)$ denotes the cumulative count of the file requests (for each (user, file) tuple), and the vector $\bm{\Theta}(t)= \big(\bm{\theta}^i(t), 1 \leq i \leq n \big)$ denotes the perturbed cumulative file request counts obtained upon adding a scaled i.i.d. Gaussian noise vector to $\bm{X}(t)$. It is also possible to sample a fresh Gaussian vector $\bm{\gamma}_t$ in step \ref{addG} at every time slot, leading to a high probability regret bound \citep{devroye2015random}. The following Theorem gives an upper bound on the regret achieved by \texttt{LeadCache}:
\begin{theorem}\label{FTPL_inelastic_th}
	The expected regret of the \texttt{LeadCache} policy is upper bounded as:
	\begin{eqnarray*}
	\mathbb{E}\big(R^{\texttt{LeadCache}}_T\big) \leq \kappa n^{3/4}d^{1/4}\sqrt{mCT},	
	\end{eqnarray*}
where $\kappa=O\big(\textsf{poly-log}(N/C)\big),$ and the expectation is taken with respect to the random noise added by the policy.
\end{theorem}
Note that in contrast with the generic regret bound of \citet{suggala2020online} for non-convex problems, our regret-bound has only logarithmic dependence on the ambient dimension $N$.  
\paragraph{Proof outline:} Our analysis of the \texttt{LeadCache} policy uses the elegant stochastic smoothing framework developed in  \citet{abernethy2016perturbation, abernethy2014online, cohen2015following, lee2018analysis}. 
See Section \ref{FTPL_inelastic_th_proof} of the supplementary material for the proof of Theorem \ref{FTPL_inelastic_th}. 
\subsection{Fast approximate implementation} \label{implementation}
The computationally intensive procedures in Algorithm \ref{uncoded_inelastic} are (I) solving the optimization problem in step \ref{algo_step} to determine the virtual caching actions and (II) translating the virtual actions back to the physical caching actions in step \eqref{algo_step2}. Since the perturbed vector $\bm{\Theta}(t)$ is obtained by adding white Gaussian noise to the cumulative request vector $\bm{X}(t)$, some of its components could be negative. For maximizing the objective \eqref{algo_step}, it is clear that if some coefficient $\theta^i_f(t)$ is negative for some $(i,f)$ tuple, it is feasible and optimal to set the  virtual action variable $z^i_f$ to zero. 
Hence, steps \eqref{algo_step} and \eqref{algo_step2} of Algorithm \ref{uncoded_inelastic} may be combined as:
\begin{eqnarray} \label{master_opt}
	\bm{y}(t) \gets \arg \max_{\bm{y} \in \mathcal{Y}} \underbrace{\sum_{i \in \mathcal{I},f \in [N]} (\theta^i_f(t))^+ \bigg(\min \big(1, \sum_{j \in \partial^+(i)} y^j_f \big)\bigg)}_{L(\bm{y})},
\end{eqnarray}
where $x^+ \equiv \max(0,x).$ 
%
%
%
Incidentally, we find that problem \eqref{master_opt} is mathematically identical to the uncoded \emph{Femtocaching} problem with known file request probabilities studied by \citet[Section III]{shanmugam2013femtocaching}. In the same paper, the authors proved the problem \eqref{master_opt} to be \textbf{NP-Hard}. The authors also proposed a complex iterative rounding method for the LP relaxation of the problem, where, in each iteration, one needs to compute certain matchings. We now propose two simple \emph{linear-time} rounding techniques that  enjoy the same approximation guarantee. 
\paragraph{LP Relaxation:}
 We now introduce a new set of variables $z^i_f:= \min(1, \sum_{j \in \partial^+(i)}y^j_f), \forall i, f,$ and relax the integrality constraints to arrive at the following LP:
\begin{eqnarray} \label{mainLP}
\max \sum_{i,f}(\theta^i_f(t))^+ z^i_f,	
\end{eqnarray}
 Subject to,
\begin{eqnarray}
 &&z^i_f \leq \sum_{j \in \partial^+(i)} y^j_f, ~~\forall i \in \mathcal{I},f \in [N]	\label{constr1}\\
&&\sum_{f=1}^N y_f^j  \leq  C, ~~ \forall j \in \mathcal{J},	\label{cap_ctr2}\\
&&0 \leq y^j_f \leq 1,~~ \forall j \in \mathcal{J}, f \in [N]; 0 \leq z^i_f \leq 1,~~ \forall i \in \mathcal{I}, f \in [N]. \label{relaxed_constr1} 
\end{eqnarray}
 Denote the objective function for the problem \eqref{master_opt} by $L(\bm{y})$ and its optimal value (over $\mathcal{Z}$) by $\textsf{OPT}.$ Let $\bm{y}^*$ be an optimal solution to the relaxed LP \eqref{mainLP} and $\mathcal{Z}_{\textrm{rel}}$ be the corresponding relaxed feasible set. Since LP \eqref{mainLP} is a relaxation to \eqref{master_opt}, it naturally holds that 
 $ L(\bm{y}^*)  \geq \textsf{OPT}.$ To round the resulting cache allocation vector $\bm{y}^*$ to an integral one, we consider the following two rounding schemes - (1) Deterministic Pipage rounding and (2) Randomized sampling-based rounding. 
 \paragraph{1. Pipage Rounding:} The general framework of \emph{Pipage rounding} was introduced by \citet{ageev2004pipage}. Our rounding technique, given in Algorithm \ref{pipage}, is markedly simpler compared to Algorithm 1 of \citet{shanmugam2013femtocaching}. While we round two fractional allocations of a single cache at a time, the rounding procedure of \citet{shanmugam2013femtocaching}  jointly rounds several allocations in multiple caches at the same time by computing matchings in a bipartite graph. 

\begin{algorithm}
\caption{Cache-wise Deterministic Pipage rounding}
\label{pipage}
\begin{algorithmic}[1]
\STATE $\bm{ y} \gets$ Solution of the $\textsf{LP}$ \eqref{mainLP}. \label{LP_sol}\
\WHILE{$\bm{ y}$ is not integral}
\STATE Select a cache $j$ with two fractional variables $y^j_{f_1}$ and $y^j_{f_2}.$ \alglinelabel{cache_sel}
\STATE Set $\epsilon_1 \gets \min (y^j_{f_1}, 1 - y^j_{f_2}), \epsilon_2 \gets \min(1-y^j_{f_1}, y^j_{f_2} ).$ \alglinelabel{dir11}
\STATE Define two new feasible cache-allocation vectors $\bm{\alpha}, \bm{\beta}$ as follows: \alglinelabel{dir13}
\begin{eqnarray}
\alpha^{j}_{f_1} \gets y^j_{f_1}-\epsilon_1, \alpha^j_{f_2}\gets y^j_{f_2}+\epsilon_1,  ~~~~ \textrm{ and } ~~~~ \alpha^k_f \gets y^k_f, \textrm{ otherwise}, \label{rd1}\\ 
\beta^j_{f_1} \gets y^j_{f_1}+\epsilon_2, \beta^j_{f_2}\gets y^j_{f_2}-\epsilon_2, ~~~~ \textrm{ and }  ~~~~ \beta^k_f \gets y^k_f, \textrm{ otherwise.}  \label{rd11}
\end{eqnarray}
\STATE Set $\bm y \gets \arg\max_{\bm{x} \in \{\bm{\alpha, \beta}\}} \phi(\bm{x}). $ \alglinelabel{rd2}
\ENDWHILE
\STATE \textbf{return} $\bm{y}.$ 
\end{algorithmic}
\end{algorithm}
\paragraph{Design:} The key to our deterministic rounding procedure is to consider the following surrogate objective function $\phi(\bm{y})$ instead of the original objective $L(\bm{y})$ as given in Eqn.\ \eqref{master_opt}:
\begin{eqnarray}\label{surr_def12}
\phi(\bm y)\equiv \sum_{i,f} (\theta^i_f(t))^+ \big(1- \prod_{j \in \partial^+(i)}(1-y^j_f) \big).
\end{eqnarray}
Following a standard algebraic argument \citep[Eqn.\ (16)]{ageev2004pipage}, we have: 
\begin{eqnarray} \label{bd_ineq}
L(\bm y)\stackrel{(a)}{\geq} \phi(\bm y) \geq \bigg(1-(1-\frac{1}{\Delta})^\Delta\bigg)L(\bm y),~~ \forall \bm{y} \in [0,1]^N,
\end{eqnarray}
where $\Delta \equiv \max_{i \in \mathcal{I}}|\partial^+(i)|.$ Note that inequality (a) holds with equality for all binary vectors $\bm{y} \in \{0,1\}^{mN} $. 
Our Pipage rounding procedure, given in Algorithm \ref{pipage}, begins with an optimal solution of the LP \eqref{mainLP}. Then it iteratively perturbs two fractional variables (if any) in a single cache in such a way that the value of the surrogate objective function $\phi(\bm{y})$ never decreases while at least one of the two fractional variables is rounded to an integer. Step \eqref{dir11} ensures that the feasibility is maintained at every step of the roundings. Upon termination (which occurs within $O(mN)$ steps), the rounding procedure yields a feasible integral allocation vector $\hat{\bm{y}}$ with an objective value $L(\hat{\bm{y}}),$ which is within a factor of $1- (1-\frac{1}{\Delta})^\Delta$ of the optimum objective. The following theorem formalizes this claim.
\begin{theorem} \label{approx_thm}
Algorithm \ref{pipage} is an $\alpha = 1-(1-\frac{1}{\Delta})^\Delta$ approximation algorithm for the problem \ref{master_opt}.	
\end{theorem}
See Section \ref{approx_thm_proof} of the supplementary material for the proof of Theorem \ref{approx_thm}. Note that the Pipage rounding procedure, although effective in practice, is not known to have a formal regret bound. 
\paragraph{2. An efficient policy for achieving a sub-linear $\alpha$-regret:}
Since the offline problem \eqref{master_opt} is \textbf{NP-Hard}, a natural follow-up problem is to design a policy with a sub-linear $\alpha$-regret. Recall that $\alpha$-regret is defined similarly as the usual static regret where the reward accrued by the offline oracle policy (\emph{i.e.,} the first term in Eqn.\ \eqref{regret_def}) is discounted by a factor of $\alpha$ \citep{kalai2005efficient}. Note that directly using the Pipage-rounded solution from Algorithm \ref{pipage} does not necessarily yield an online policy with a sub-linear $\alpha$-regret
\citep{kakade2009playing, garber2021efficient, hazan2018online, fujita2013combinatorial}. 
In the following, we give an efficient offline-to-online reduction that makes only a \emph{single} query to a linear-time randomized rounding procedure per iteration. In our reduction, the following notion of an $\alpha$ point-wise approximation algorithm \citep{kalai2005efficient} plays a pivotal role.
\begin{definition}[$\alpha$ point-wise approximation]
	 For a  feasible set $\mathcal{Z}$ in the non-negative orthant and a non-negative input vector $\bm{x},$ consider the Integer Linear Program $\max_{\bm{z} \in \mathcal{Z}}  \bm{z} \cdot \bm{x}.$
	 Let $\mathcal{Z}_{\textrm{rel} } \supseteq \mathcal{Z}$ be a relaxation of the feasible set $\mathcal{Z}$ and let $\bm{z} \in \arg\max_{\bm{z} \in \mathcal{Z}_{\textrm{rel}}} \bm{z} \cdot \bm{x}$ be an optimal solution of the relaxed ILP. If for some $\alpha > 0,$ a (randomized) rounding algorithm $A$ returns a feasible solution $\hat{\bm{z}} \in \mathcal{Z}$ such that $ \mathbb{E}\hat{z}_i \geq \alpha z_i, \forall i$ and for any input $\bm{x}$, we call the algorithm $A$ an $\alpha$ point-wise approximation. 
\end{definition}
It immediately follows that for any $\alpha$ point-wise approximation algorithm for the problem \ref{mainLP}, if $\bm{z}_t \in \mathcal{Z}_{\textrm{rel}}$ be the relaxed virtual action at time $t$, we have 
$\sum_{t=1}^T \mathbb{E}\big[\bm{\hat{z}}_t\big] \cdot \bm{x}_t \geq 	\alpha \sum_{t=1}^T \bm{z}_t \cdot \bm{x}_t$,
where the inequality follows from the point-wise approximation property. Thus, for any $\bm{z}^* \in \mathcal{Z},$ the $\alpha$-regret of the virtual policy may be upper bounded as
\begin{eqnarray} \label{alpha-guarantee}
	\alpha \sum_{t=1}^T \bm{x}_t \cdot \bm{z}^* - \sum_{t=1}^T \mathbb{E}\big[\bm{\hat{z}}_t\big] \cdot \bm{x}_t \leq \alpha \big(\sum_{t=1}^T \bm{x}_t \cdot \bm{z}^* - \sum_{t=1}^T \bm{x}_t \cdot \bm{z}_t  \big) \stackrel{(b)}{\leq} \alpha \mathbb{E}(\tilde{R}_T^{\texttt{LeadCache}}),
\end{eqnarray}
 where $\tilde{R}_T^{\texttt{LeadCache}}$ is an upper bound to the regret of the \texttt{LeadCache} policy with the relaxed actions given by the solution of the LP \eqref{mainLP}. We bound the quantity $\mathbb{E}(\tilde{R}_T^{\texttt{LeadCache}})$ in the following Proposition.
\begin{proposition}\label{relaxed-regret-bound}
For an appropriate learning rate sequence $\{\eta_t\}_{t \geq 1}$, the expected regret of the \textrm{LeadCache} policy with the relaxed action set $\mathcal{Z}_{\textrm{rel}}$ can be bounded as follows:
	\[ \mathbb{E}(\tilde{R}_T^{\texttt{LeadCache}}) \leq \kappa_1 n^{3/4} \sqrt{dmCT}, \]
	where $\kappa_1$ is poly-logarithmic in $N$ and $n$. 
\end{proposition}
 See Section \ref{relaxed-regret-bound-proof} of the supplementary materials for a proof sketch of the above result. 
  In the following, we design an $\alpha$ point-wise approximate randomized rounding scheme.
\paragraph{Randomized Rounding via Madow's sampling:}
The key ingredient to our $\alpha$ point-wise approximation oracle is Madow's systematic sampling scheme taken from the statistical sampling literature \citep{madow1949theory}. For a set of \emph{feasible} inclusion probabilities $\bm{p}$ on a set of items $[N]$, Madow's scheme outputs a subset $S$ of size $C$ such that the $i$\textsuperscript{th} element is included in the subset $S$ with probability $p_i, 1\leq i \leq N.$ For this sampling scheme to work, it is necessary and sufficient that the inclusion probability vector $\bm{p}$ satisfies the following feasibility constraint: 
\begin{eqnarray}\label{feasibility-constraint}
	\sum_{i=1}^N p_i =C, \textrm{ and }
	0\leq p_i \leq 1, \forall i \in [N]. 
\end{eqnarray}
The pseudocode for Madow's sampling is given in Section \ref{madow-pseudocode} in the supplement.
%
Our proposed $\alpha$ point-wise approximate rounding scheme independently samples $C$ files in each cache in accordance with the inclusion probability given by the fractional allocation vector $\bm{y}$ obtained from the solution of the relaxed LP \eqref{mainLP}. From the constraints of the LP, it immediately follows that the inclusion vector $\bm{y}^j$ satisfies the feasibility constraint \eqref{feasibility-constraint}; hence the above process is sound. The overall rounding scheme is summarized in Algorithm \ref{alpha-rounding}.  
To show that the resulting rounding scheme satisfies the $\alpha$ point-wise approximation property, note that for all $i \in \mathcal{I}$ and $f \in [N]:$
\begin{eqnarray*}
\mathbb{P}(\hat{z}^i_f =1) = \mathbb{P}( \bigvee_{j \in \partial^+(i)} \hat{y}^j_f =1)
 \stackrel{(a)}{=} 1- \prod_{j \in \partial^+(i)}(1-y^j_f) 
 \stackrel{(b)}{\geq} 1 - e^{-\sum_{j \in \partial^+(i)} y^j_f } 
 \stackrel{(c)}{\geq} 1 - e^{-z^i_f} 
 \stackrel{(d)}{\geq} (1-\frac{1}{e})z^i_f, 	
\end{eqnarray*}
where the equality (a) follows from the fact that rounding in each caches are done independently of each other, the inequality (b) follows from the standard inequality $\exp(x) \geq 1+x, \forall x \in \mathbb{R}$,  the inequality (c) follows from the feasibility constraint \eqref{constr1} of the LP, and finally the inequality (d) follows from the concavity of the function $1-\exp(-x)$ and the fact that $0\leq z^i_f \leq 1.$ Since $\hat{z}^i_f$ is binary, it immediately follows that the randomized rounding scheme in Algorithm \ref{alpha-rounding} is an $\alpha$ point-wise approximation with $\alpha = 1-\nicefrac{1}{e}.$ We formally state the result in the following Theorem.
\begin{theorem} \label{alpha-regret-th}
The \texttt{LeadCache} policy, in conjunction with the randomized rounding scheme with Madow's sampling (Algorithm \ref{alpha-rounding}), achieves an $\alpha = 1-e^{-1}$-regret bounded by $\tilde{O}(n^{3/4}\sqrt{dmCT}).$	
\end{theorem}

\begin{algorithm} 
\caption{Randomized Rounding with Madow's Sampling Scheme}
\label{alpha-rounding}
\begin{algorithmic}[1]
\REQUIRE Fractional cache allocation $(\bm{y, z}).$
\ENSURE A rounded allocation $(\bm{\hat{y}, \hat{z}})$
\STATE 	Let $\bm{y}$ be the output of the LP \eqref{mainLP}.  
\FOR {each cache $j \in \mathcal{J}$}
\STATE Independently sample $S_j$ a set of $C$ files with the probability vector $\bm{y}^j$ using Madow's sampling scheme \ref{madow}.
\STATE $\hat{y}^j_f \gets 1(f \in S_j)$.
\ENDFOR 
\STATE $\hat{z}^i_f \gets \bigvee_{j \in \partial^+(i)} \hat{y}^j_f, \forall i \in \mathcal{I}.$
\STATE \textbf{Return} $(\bm{\hat{y}, \hat{z}})$
\end{algorithmic}
\end{algorithm}


\section{Bounding the Number of Fetches} \label{bounded_download}
Fetching files from the remote server to the local caches consumes bandwidth and increases network congestion. 
Under non-prefetching policies, such as \textsf{LRU}, \textsf{FIFO}, and \textsf{LFU}, a file is fetched if and only if there is a cache miss. 
Hence, for these policies, it is enough to bound the cache miss rates in order to control the download rate.
 However, since the \texttt{LeadCache} policy decouples the fetching process from the cache misses, in addition to a small regret, we need to ensure that the number of file fetches remains small as well. 
We now prove the surprising result that if the file request process satisfies a mild regularity property, the file fetches \emph{stop almost surely after a finite time}. 
Note that our result is of a different flavor from the long line of work that minimizes the switching regret under adversarial inputs but yield a much weaker bound on the number of fetches \citep{SwRegret, devroye2015random, kalai2005efficient, geulen2010regret}. 
 
\textbf{A. Stochastic Regularity Assumption:} Let $\{\bm{X}(t)\}_{t \geq 1}$ be the cumulative request-arrival process. We assume that, there exists a set of non-negative numbers $\{p^i_f\}_{i\in \mathcal{I}, f\in [N]}$ such that for any $\epsilon>0:$
\begin{eqnarray} \label{renewal_conc}
\sum_{t=1}^\infty \sum_{i \in \mathcal{I}, f \in [N]}\mathbb{P}\big( \big|\frac{\bm{X}^i_f(t)}{t}- p^{i}_f\big| \geq \epsilon\big) < \infty.
\end{eqnarray}
Using the first Borel-Cantelli Lemma, the regularity assumption \textbf{A} implies that the process $\{\bm{X}(t)\}_{t\geq 1}$ satisfies the strong-law:
$ \nicefrac{\bm{X}^i_f (t)}{t} \to p^i_f, \textrm{a.s.}, \forall i \in \mathcal{I},f \in [N]. $
However, the converse may not be true. Nevertheless,  
 the assumption \textbf{A} is quite mild and holds, \emph{e.g.}, when the file request sequence is generated by a renewal process having an inter-arrival distribution with a finite fourth moment (See Section \ref{renewal_proc} of the supplementary material for the proof). Define a fetch event $F(t)$ to take place at time slot $t$ if the cache configuration at time slot $t$ is different from that of at time slot $t-1,$ \emph{i.e.,} $F(t) := \{ \bm{y}(t) \neq \bm{y}(t-1)\}.$ The following is our main result in this section:
\begin{theorem} \label{fetch}
	Under the stochastic regularity assumption \textbf{A}, the file fetches to the caches stop after a finite time with probability $1$ under the \texttt{LeadCache} policy. 
\end{theorem}
Please refer to Section \ref{fetch_proof} of the supplementary material for the proof of Theorem \ref{fetch}. This Theorem implicitly assumes that the optimization problems in steps 7 and 8 are solved exactly at every time.

\section{A Minimax Lower Bound} \label{lower_bound}
In this section, we establish a minimax lower bound to the regret for the \textsf{Bipartite Caching} problem. Recall that our reward function \eqref{reward_definition} is non-linear. 
As such, the standard lower bounds, such as Theorem $5.1$ and $5.12$ of \citet{orabona2019modern}, Theorem $5$ of  \citet{abernethy2008optimal} are insufficient for our purpose. 
The first regret lower bound for the \textsf{Bipartite Caching} problem was given by \citet{SIGMETRICS20}. We now strengthen this bound by utilizing results from graph coloring.
   
\begin{theorem}[Regret Lower Bound] \label{lb_thm}
	For a catalog of size $N \geq \max( 2 \frac{d^2Cm}{n}, 2mC)$ the regret of any online caching policy $\pi$ is lower bounded as: 
	$ R_T^\pi \geq \max ( \sqrt{\frac{mnCT}{2\pi }}, d \sqrt{\frac{mCT}{2\pi}})  - \Theta(\frac{1}{\sqrt{T}}).$
\end{theorem}	
\paragraph{Proof outline:} 
We use the standard probabilistic method where the worst-case regret is lower bounded by the average regret over an ensemble of problems. 
 To lower-bound the cumulative reward accrued by the optimal offline policy,
 we construct an offline static caching configuration, where each user sees different files on each of the connected caches (\emph{local exclusivity}). 
 This construction effectively linearizes the reward function that can be analyzed using the probabilistic \emph{Balls-into-Bins} framework and graph coloring theory.
 See Section \ref{lower_bound_proof} of the supplementary material for the proof. 
\vspace{-10pt}
\paragraph{Approximation guarantee for \texttt{LeadCache}:} Theorem \ref{FTPL_inelastic_th} and Theorem \ref{lb_thm}, taken together, imply that the \texttt{LeadCache} policy achieves the optimal regret within a factor of $\tilde{O}(\min ( (nd)^{1/4}, (\nicefrac{n}{d})^{3/4} )).$ 
Hence, irrespective of $d$, the \texttt{LeadCache} policy is regret-optimal up to a factor of $\tilde{O}(n^{3/8}).$ 

\section{Experiments} \label{experiments}
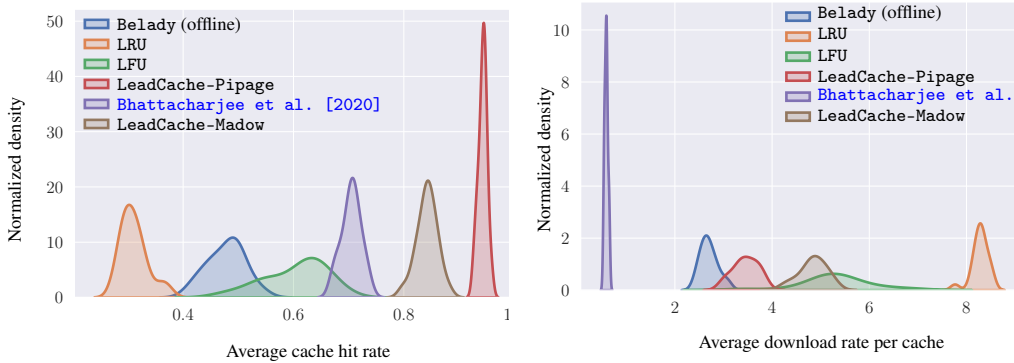
\begin{figure}
\centering
\begin{minipage}{0.43\textwidth}
   \hspace{-20pt}
         \begin{overpic}[height=2in]{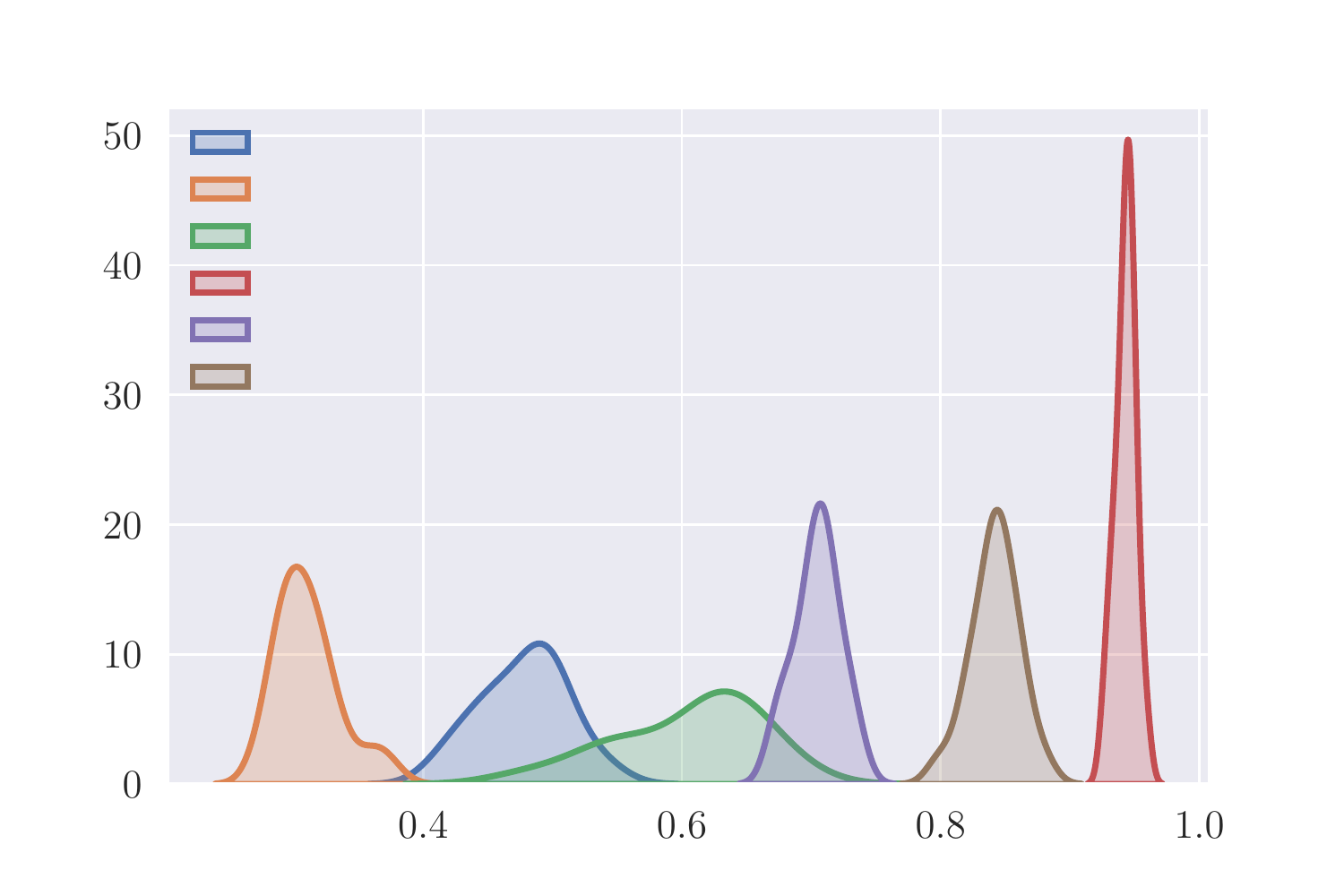}
         \put(20,55){\scriptsize{\texttt{Belady} (offline)}}
         \put(20,51.5){\scriptsize{\texttt{LRU}}}
         \put(20,48){\scriptsize{\texttt{LFU}}}
         \put(20,44.5){\scriptsize{\texttt{LeadCache-Pipage}}}
          \put(20,41){\scriptsize{\texttt{\textcolor{blue}{Bhattacharjee et al.\ [2020]}}}}
         \put(20, 37.5){\scriptsize{\texttt{LeadCache-Madow}}}
         \put(1, 18){\rotatebox{90}{\scriptsize{Normalized density}}}
         \put(39,-2){\scriptsize{Average cache hit rate}}
          \end{overpic}
 \end{minipage}
 \begin{minipage}{0.43\textwidth}
         \begin{overpic}[height=2in]{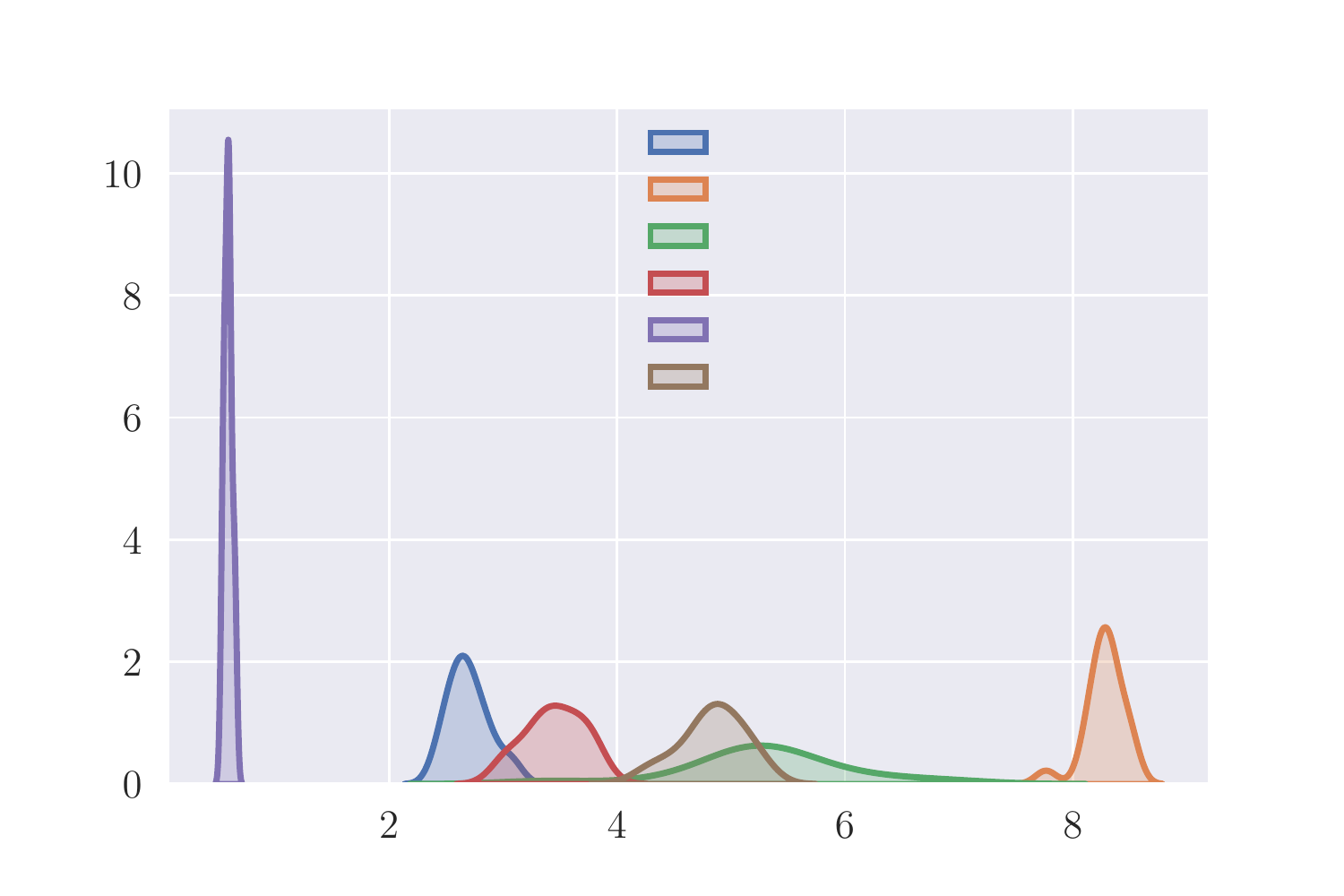}
         \end{overpic}
         \put(-100,121){\scriptsize{\texttt{Belady} (offline)}}
         \put(-100,113){\scriptsize{\texttt{LRU}}}
         \put(-100,105){\scriptsize{\texttt{LFU}}}
         \put(-100,97){\scriptsize{\texttt{LeadCache-Pipage}}}
         \put(-100,90){\scriptsize{\texttt{\textcolor{blue}{Bhattacharjee et al.}}}}
         \put(-100, 82){\scriptsize{\texttt{LeadCache-Madow}}}
         \put(-206, 40){\rotatebox{90}{\scriptsize{Normalized density}}}
         \put(-145,-4){\scriptsize{Average download rate per cache}}
    \end{minipage}
    \caption{\small{Empirical distributions of (a) Cache hit rates and (b) Fetch rates of different caching policies.}}
    \label{hit-download-rates}
\end{figure}

In this section, we compare the performance of the \texttt{LeadCache} policy with standard caching policies.
\paragraph{Baseline policies:} Under the  \textsf{LRU} policy \citep{borodin2005online}, each cache considers the set of all requested files from its connected users independently of other caches. In the case of a \emph{cache-miss}, the \textsf{LRU} policy fetches the requested file into the cache while evicting a file that was requested \emph{least recently}. The \textsf{LFU} policy works similarly to the \textsf{LRU} policy with the exception that, in the case of a cache-miss, the policy evicts a file that was requested the \emph{least number of times} among all files currently on the cache. Finally, for the purpose of benchmarking, we also experiment with Belady's offline \textsf{MIN} algorithm \citep{aho1971principles}, which is optimal in the class of non-pre-fetching reactive policies for each \emph{individual} cache when the entire file request sequence is known a priori. Note that Belady's algorithm is \emph{not} optimal in the network caching setting as it does not consider the adjacency relations between the caches and the users. 
The heuristic multi-cache policy by \citet{SIGMETRICS20} uses an \textsf{FTPL} strategy for file prefetching while approximating the reward function \eqref{reward_definition} by a linear function. 
\paragraph{Experimental Setup:} In our experiments, we use a publicly available anonymized production trace from a large CDN provider available under a BSD 2-Clause License \citep{berger2018practical, trace}. The trace data consists of three fields, namely, request number, file-id, and file size. In our experiments, we construct a random Bipartite caching network with $n=30$ users and $m=10$ caches. Each cache is connected to $d=8$ randomly chosen users. Thus, every user is connected to $\approx 2.67$ caches on average.
The storage capacity $C$ of each cache is taken to be $10\%$ of the catalog size. 
We divide the trace consisting of the first $\sim 375K$ requests into $20$ consecutive sub-intervals. 
File requests are assigned to the users sequentially before running the experiments on each of the sub-intervals. The code for the experiments is available online \citep{LeadCacheCode}.
%
%
%
\paragraph{Results and Discussion:} Figure \ref{hit-download-rates} shows the performance of different caching policies in terms of the average cache hit rate per file and the average number of file fetches per cache. 
From the plots, it is clear that the \texttt{LeadCache} policy, with any of its Pipage and Madow rounding variants, outperforms the baseline policies in terms of the cache hit rate. Furthermore, we find that the deterministic Pipage rounding variant empirically outperforms the randomized Madow rounding variant (that has a guaranteed sublinear $\alpha$-regret) in terms of both hit rate and fetch rates. On the flip side, the heuristic policy of \citet{SIGMETRICS20} incurs a fewer number of file fetches compared to the \texttt{LeadCache} policy. 
From the plots, it is clear that the \texttt{LeadCache} policy excels by effectively coordinating the caching decisions among different caches and quickly adapting to the dynamic file request pattern. Section \ref{addl_plots} of the supplementary material gives additional plots for the popularity distribution and the temporal dynamics of the policies for a given file request sequence.
 
\section{Conclusion and Future Work} \label{conclusion}
In this paper, we proposed an efficient network caching policy called \texttt{LeadCache}. We showed that the policy is competitive with the state-of-the-art caching policies, both theoretically and empirically. We proved a lower bound for the achievable regret and established that the number of file-fetches incurred by our policy is finite under reasonable assumptions. 
Note that \texttt{LeadCache} optimizes the cumulative cache hits without considering fairness among the users. In particular, \texttt{LeadCache} could potentially ignore a small subset of users who request unpopular content. A future research direction could be to incorporate fairness into the \texttt{LeadCache} policy so that \emph{each} user incurs a small regret \citep{destounis2017alpha}. Finally, it will be interesting to design caching policies that enjoy strong guarantees for the regret and the competitive ratio simultaneously \citep{daniely2019competitive}. 
\section{Acknowledgement} \label{ack}
This work was partially supported by the grant IND-417880 from Qualcomm, USA, and a research grant from the Govt.\ of India  under the IoE initiative. The computational  work reported on this paper was performed on the AQUA Cluster at
the High Performance Computing Environment of IIT Madras. 
The authors would also like to thank Krishnakumar from IIT Madras for his help with a few illustrations appearing on the paper. 
\clearpage
\bibliography{OCO}
\bibliographystyle{unsrtnat}
\clearpage
\onecolumn
\section{Supplementary Material for the paper \emph{\texttt{LeadCache}: Regret-Optimal Caching in Networks} by Debjit Paria and Abhishek Sinha} \label{appendix}

\subsection{Proof of Lemma \ref{regret_comp}} \label{regret_comp_proof}
From equation \eqref{virtual_physical}, we have that for any file request vector $\bm{x}$ and virtual caching configuration $\bm{z}:$ 
\begin{eqnarray*}
r(\bm{x}, \bm{z}) = \langle \bm{x}, \bm{z} \rangle \leq q(\bm{x}, \psi(\bm{z})).	
\end{eqnarray*}
Thus for any file request sequence $\{\bm{x}_t\}_{t \geq 1},$ we have: 
\begin{eqnarray} \label{tot_rew_bd}
\sum_{t=1}^T r(\bm{x}_t, \bm{z}_t) \leq \sum_{t=1}^T q(\bm{x}_t, \psi(\bm{z}_t)).	
\end{eqnarray}
On the other hand, let $\bm{y}_* \in \arg \max_{\bm{y} \in \mathcal{Y}} \sum_{t=1}^Tq(\bm{x}_t, \bm{y})$ be an optimal static offline cache configuration vector corresponding to the file requests $\{\bm{x}_t\}_{t=1}^T.$ Consider a candidate static virtual cache configuration vector $\bm{z}_* \in \mathcal{Z}$ defined as:
\begin{eqnarray*}
\bm{z}_*^i \equiv \min\bigg\{ \bm{1}_{N\times 1}, \big(\sum_{j \in \partial^+(i)}\bm{y}_*^j\big) \bigg\}, ~~1 \leq i \leq n.
\end{eqnarray*}
We have
\begin{eqnarray} \label{static_opt}
\max_{\bm{z} \in \mathcal{Z}} \sum_{t=1}^T r(\bm{x}_t, \bm{z})	\geq   \langle \sum_{t=1}^T \bm{x}_t, \bm{z}_* \rangle = \max_{\bm{y} \in \mathcal{Y}} \sum_{t=1}^T q(\bm{x}_t, \bm{y}).
\end{eqnarray}

Combining Eqns. \eqref{tot_rew_bd} and \eqref{static_opt}, we conclude that
\begin{eqnarray} \label{reg_ineq1}
\max_{\bm{y} \in \mathcal{Y}} \sum_{t=1}^T q(\bm{x}_t, \bm{y}) - \sum_{t=1}^T q(\bm{x}_t, \psi(\bm{z}_t)) \leq \max_{\bm{z} \in \mathcal{Z}} \sum_{t=1}^T r(\bm{x}_t, \bm{z}) - \sum_{t=1}^T r(\bm{x}_t, \bm{z}_t).
\end{eqnarray}
Taking supremum of both sides of inequality \eqref{reg_ineq1} over all possible file request sequences $\{\bm{x}_t\}_{t \geq 1}$ yields the result.
$\hfill \blacksquare$

%
%

\subsection{Proof of Theorem \ref{FTPL_inelastic_th}} \label{FTPL_inelastic_th_proof}
Keeping Lemma \ref{regret_comp} in view, to prove the desired regret upper bound for the \texttt{LeadCache} policy, it is enough to bound the regret for the virtual policy $\pi^{\textrm{virtual}}$ only.
Following \cite{cohen2015following} we derive a general expression for the regret upper bound applicable to any linear reward function under an anytime \textsf{FTPL} policy. This is accomplished in the following steps. First, we extend the argument of \cite{cohen2015following} to the anytime setting.  
Then, we specialize this bound to our problem setting. 

Recall the notations used in the paper - the aggregate file-request sequence from all users is denoted by $\{\bm{x}_t\}_{t \geq 1}$ and the virtual cache configuration sequence is denoted by $\{\bm{z}_t\}_{t \geq 1}$. Define the cumulative requests up to time $t$ as:
\[\bm{X}_t = \sum_{\tau=1}^{t-1} \bm{x}_\tau.\]
Note that the \texttt{LeadCache} policy chooses the virtual cache configuration at time slot $t$ by solving the following optimization problem at time slot $t$:
\begin{eqnarray}\label{opt_z}
	\bm{z}_t = \arg\max_{\bm{z} \in \mathcal{Z}} \langle \bm{z}, \bm{X}_t + \eta_t \bm{\gamma}\rangle, 
\end{eqnarray}
 where each of the $Nn$ components of the random vector $\bm{\gamma}$ is sampled independently from the standard Gaussian distribution, and $\mathcal{Z}$ denotes the set of all feasible virtual cache configurations as defined earlier in the paper. Next, we define the following potential function: 
 \begin{eqnarray}\label{potential_function_def}
 \Phi_{\eta_t}(\bm{x}) \equiv \mathbb{E}_{\bm{\gamma}} \bigg[\max_{\bm z \in \bm{\mathcal Z}}\langle \bm{z}, \bm{x}+ \eta_t \bm{\gamma} \rangle \bigg].
 \end{eqnarray}
 Since the perturbation r.v.\ $\bm{\gamma}$ is Gaussian, it follows that the potential function $\phi_{\eta_t} (\bm{x})$ is twice continuously differentiable \citep[Lemma 1.5]{abernethy2016perturbation}.   
 Furthermore, since the max function is convex, we may interchange the expectation and gradient to obtain 
 $\grad \Phi_{\eta_t}(\bm{X}_t) = \mathbb{E}(\bm{z}_t)$ \citep[Proposition 2.2]{bertsekas1973stochastic}. Thus we have: 
 \begin{eqnarray}\label{Th6eq1}
 \langle \grad \Phi_{\eta_t}(\bm{X}_t), \bm{x}_t\rangle = \mathbb{E}\langle \bm{z}_t, \bm{x}_t \rangle. 	
 \end{eqnarray}
 To upper bound the regret of the \texttt{LeadCache} policy, we expand $\Phi_{\eta_{t}}(\bm{X}_{t+1})$ in second-order Taylor's series as follows:
 \begin{eqnarray} \label{taylor1}
 &&\Phi_{\eta_{t}}(\bm{X}_{t+1}) \nonumber \\
  &=& \Phi_{\eta_{t}}(\bm{X}_t+\bm{x}_t) \nonumber \\
 &=& \Phi_{\eta_t}(\bm{X}_t) + \langle \grad \Phi_{\eta_t}(\bm{X}_t), \bm{x}_t\rangle + \frac{1}{2}\bm{x}_t^T\grad^2\Phi_{\eta_t}(\tilde{\bm X}_t)\bm{x}_t,
 \end{eqnarray}
 where $\tilde{\bm X}_t$ lies on the line segment joining $\bm{X}_t$ and $\bm{X}_{t+1}.$
 Plugging in the expression of the inner product from Eqn.\ \eqref{Th6eq1} in expression \eqref{taylor1}, we obtain:
 \begin{eqnarray}\label{expected_reward}
 	 \mathbb{E}\langle \bm{z}_t, \bm{x}_t \rangle =  \Phi_{\eta_t}(\bm{X}_{t+1}) - \Phi_{\eta_t}(\bm{X}_t)  - \frac{1}{2}\bm{x}_t^T\grad^2\Phi_{\eta_t}(\tilde{\bm X}_t)\bm{x}_t.
 \end{eqnarray}
 Summing up Eqn.\ \eqref{expected_reward} from $t=1$ to $T$, the total expected reward accrued by the \texttt{LeadCache} policy may be computed to be: 
 \begin{eqnarray*}
 	&&\mathbb{E}\big(Q^{\texttt{LeadCache}}(T)\big) \nonumber \\
 	&=& \sum_{t=1}^T\mathbb{E}\langle \bm{z}_t, \bm{x}_t \rangle \nonumber \\
 	 &=& \sum_{t=1}^T\bigg(\Phi_{\eta_t}(\bm{X}_{t+1})- \Phi_{\eta_t}(\bm{X}_t) \bigg)- \frac{1}{2}\sum_{t=1}^T \bm{x}_t^T\grad^2\Phi_{\eta_t}(\tilde{\bm X}_t)\bm{x}_t \nonumber \\
 	 &=& \sum_{t=1}^T\bigg(\Phi_{\eta_t}(\bm{X}_{t+1})-\Phi_{\eta_{t+1}}(\bm{X}_{t+1})+\Phi_{\eta_{t+1}}(\bm{X}_{t+1})- \Phi_{\eta_t}(\bm{X}_t) \bigg)- \frac{1}{2}\sum_{t=1}^T \bm{x}_t^T\grad^2\Phi_{\eta_t}(\tilde{\bm X}_t)\bm{x}_t \nonumber \\
 	 &=& \sum_{t=1}^T\bigg(\Phi_{\eta_t}(\bm{X}_{t+1})-\Phi_{\eta_{t+1}}(\bm{X}_{t+1})\bigg) + \Phi_{\eta_{T+1}}(\bm{X}_{T+1})- \Phi_{\eta_1}(\bm{X}_1) - \frac{1}{2}\sum_{t=1}^T \bm{x}_t^T\grad^2\Phi_{\eta_t}(\tilde{\bm X}_t)\bm{x}_t. \nonumber \\
 	\end{eqnarray*}
 	Next, note that 
 	\begin{eqnarray*}
 	\Phi_{\eta_{T+1}}(\bm{X}_{T+1})&=&\mathbb{E}_{\bm \gamma} \big[\max_{\bm{z} \in \mathcal{Z}} \langle \bm{z}, \bm{X}_{T+1} + \eta_{T+1} \bm{\gamma} \rangle \big] \\
 	&\stackrel{\textrm{(Jensen's ineq.\ )}}{\geq}& \max_{\bm z \in \mathcal{Z}} \big[ \mathbb{E}_{\bm \gamma}  \langle \bm{z}, \bm{X}_{T+1} + \eta_{T+1} \bm{\gamma} \rangle   \big]\\
 	&=& \max_{\bm z \in \mathcal{Z}} \langle \bm{z}, \bm{X}_{T+1} \rangle\\
 	&=& Q^*(T), 
 	 \end{eqnarray*}
 	 where recall that $Q^*(T)$ denotes the optimal cumulative reward up to time $T$ obtained by the best static policy in hindsight. Hence, from the above, we can upper bound the expected regret \eqref{regret_def} of the \texttt{LeadCache} policy as: 
 	 \begin{eqnarray}\label{FTPL_gen_bd}
 	 &&\mathbb{E}(R_T^{\texttt{LeadCache}}) \nonumber \\
 	 &=& Q^*(T)- \mathbb{E}\big(Q^{\texttt{LeadCache}}(T)\big) \nonumber \\
 	 &\leq &  \Phi_{\eta_1}(\bm{X}_1) +\underbrace{\sum_{t=1}^T\bigg(\Phi_{\eta_{t+1}}(\bm{X}_{t+1})- \Phi_{\eta_t}(\bm{X}_{t+1})\bigg)}_{(a)} +  \frac{1}{2}\sum_{t=1}^T \bm{x}_t^T\grad^2\Phi_{\eta_t}(\tilde{\bm X}_t)\bm{x}_t.  	
 	 \end{eqnarray}

\paragraph{Bounding the term (a):}
Next, to upper bound the expected regret, we control term (a) in inequality \eqref{FTPL_gen_bd}. From Eqns.\ \eqref{opt_z} and \eqref{potential_function_def}, we can write:
 \begin{eqnarray*}
 \Phi_{\eta_{t+1}}(\bm{X}_{t+1}) = \mathbb{E}\big[\langle \bm{z}_{t+1}, \bm{X}_{t+1} + \eta_{t+1}\bm{\gamma} \rangle \big],	
 \end{eqnarray*}
and
\begin{eqnarray*}
\Phi_{\eta_t}(\bm{X}_{t+1}) \geq \mathbb{E}\big[\langle \bm{z}_{t+1}, \bm{X}_{t+1} + \eta_t \bm{\gamma} \rangle \big].	
\end{eqnarray*}
Hence, each term in the summation (a) may be upper bounded as follows:
\begin{eqnarray*}
	\Phi_{\eta_{t+1}}(\bm{X}_{t+1})- \Phi_{\eta_t}(\bm{X}_{t+1}) &\leq& \mathbb{E}\big[\langle \bm{z}_{t+1}, \bm{X}_{t+1} + \eta_{t+1}\bm{\gamma} \rangle \big] - \mathbb{E}\big[\langle \bm{z}_{t+1}, \bm{X}_{t+1} + \eta_t \bm{\gamma} \rangle \big]\\
	&=& \mathbb{E} \big[ \langle \bm{z}_{t+1}, (\eta_{t+1}-\eta_t)\bm{\gamma})\rangle \big]\\
	&=& (\eta_{t+1}- \eta_t) \mathbb{E}\big[ \langle \bm{z}_{t+1}, \bm{\gamma} \rangle \big]\\
	&\leq & (\eta_{t+1}- \eta_t)\mathbb{E}\big [ \max_{z \in \mathcal{Z}}\langle \bm{z}, \bm{\gamma} \rangle \big] \\
	&=& (\eta_{t+1}-\eta_t) \mathcal{G}(\mathcal{Z}),
\end{eqnarray*}
where the quantity $\mathcal{G}(\mathcal{Z})$ is known as the \emph{Gaussian complexity} of the set $\mathcal{Z}$ of virtual configurations \cite{wainwright2019high}. Since the Gaussian perturbation $\bm{\gamma}$ has zero mean, $\mathcal{G}(\mathcal{Z})$ is non-negative due to Jensen's inequality. Substituting the above upper bound back in Eqn.\ \eqref{FTPL_gen_bd}, we notice that the summation in part (a) telescopes, yielding the following bound for the expected regret: 
\begin{eqnarray} \label{FTPL_regret_final}
\mathbb{E}(R_T^{\texttt{LeadCache}}) \leq \eta_{T+1} \underbrace{\mathcal{G}(\mathcal{Z})}_{(b)} + \frac{1}{2}\sum_{t=1}^T \underbrace{\bm{x}_t^T\grad^2\Phi_{\eta_t}(\tilde{\bm X}_t)\bm{x}_t}_{(c)}.	
\end{eqnarray}
We now upper bound each of the terms (b) and (c) as defined in the above regret bound.


 	 \paragraph{Bounding term (b) in Eqn.\ \eqref{FTPL_regret_final}:}	
 In the following, we upper bound the Gaussian complexity of the set $\mathcal{Z}$: 
\begin{eqnarray*}
\mathcal{G}(\mathcal{Z}) \equiv \mathbb{E}_{\bm \gamma} \big[ \max_{\bm z \in \mathcal{Z}} \langle \bm{z}, \bm{\gamma} \rangle \big]. 
\end{eqnarray*}
 From equation \eqref{virtual_physical}, we have for any feasible $\bm{z} \in \mathcal{Z}$:
\begin{eqnarray}
\sum_{i \in \mathcal{I}, f \in [N]}	\bm{z}^i_{f} &\leq& \sum_{i \in \mathcal{I}, f \in [N]} \sum_{ j \in \partial^+(i)} \bm{y}^j_{f}\nonumber \\
&=& \sum_{j \in \mathcal{J}} \sum_{f \in [N]} \sum_{i \in \partial^{-}(j)}\bm{y}^j_{f} \nonumber \\
&\stackrel{(a)}{\leq} & d \sum_{j \in \mathcal{J}} \sum_{f \in [N]} \bm{y}^j_{f}\nonumber \\
& \stackrel{(b)}{\leq}& dmC. \label{FTPL_pf2}
\end{eqnarray}
where, in step (a), we have used our assumption that the right-degree of the bipartite graph $\mathcal{G}$ is upper bounded by $d$, and in (b), we have used the fact that the capacity of each cache is bounded by $C$.

For any fixed $\bm{z} \in \mathcal{Z}$, the random inner-product $\langle \bm{z}, \bm{\gamma} \rangle $ follows a normal distribution with mean zero and variance $\sigma^2$ where \[\sigma^2 \equiv \mathbb{E}\langle \bm{z}, \bm{\gamma} \rangle^2 \stackrel{(a)}{=} \sum_{i \in \mathcal{I},f \in [N]} (\bm{z}^i_f)^2\stackrel{(b)}{=}\sum_{i \in \mathcal{I},f \in [N]} \bm{z}^i_f \stackrel{(c)}{\leq} dm C. \]
In the above, equality (a) follows from the fact that $\bm{\gamma}$ is a standard normal r.v., equality (b) follows from the fact that the components $z^i_f$'s are binary-valued (hence, $(z^i_f)^2=z^i_f$), and equality (c) follows from the upper bound given in Eqn.\ \eqref{FTPL_pf2}. 

Next, observe that since the feasible set $\mathcal{Z}$ is downward closed, if $\bm{z}_* \in \arg \max_{\bm{z} \in \mathcal{Z}} \langle \bm{z}, \bm{\gamma} \rangle,$ then $\gamma^i_f <0 $ implies $z^i_{*f}=0.$ Hence, we can simplify the expression for the Gaussian complexity of the set $\mathcal{Z}$ as 
\begin{eqnarray*}
\mathcal{G}(\mathcal{Z})\equiv\mathbb{E}_{\bm \gamma} \big[ \max_{\bm z \in \mathcal{Z}} \langle \bm{z}, \bm{\gamma} \rangle \big] = \mathbb{E}_{\bm{\gamma}}\big[ \max_{\bm{z} \in \mathcal{Z}} \sum_{(i,f): \gamma^i_f > 0}z^i_f \gamma^i_f\big]. 	
\end{eqnarray*}
Since all coefficients $\gamma^i_f$ in the above summation are positive, we conclude that there exists an optimal vector $\bm{z}_* \in \mathcal{Z}$ such that the inequality in Eqn.\ \eqref{virtual_physical} is met with an equality for other components of $\bm{z}$, \emph{i.e.,} $\forall (i,f): \gamma^i_f >0$, we have
\begin{eqnarray} \label{equality-virtual}
 \bm{z}^i_{*f} = \min (1, \sum_{j \in \partial^+(i)} y^j_{*f}),
 \end{eqnarray} 
for some $\bm{y}_* \in \mathcal{Y}. $ Let $\mathcal{Z}_*$ be the set of all feasible virtual caching vectors satisfying \eqref{equality-virtual} for some feasible $\bm{y}_* \in \mathcal{Y}$. 
Since the optimal virtual caching vector $\bm{z}_* \in \mathcal{Z}_*$ is completely determined by the corresponding physical caching vector $\bm{y}_* \in \mathcal{Y},$
we have that $|\mathcal{Z}_*| \leq |\mathcal{Y}|.$ Furthermore, since any of the $m$ caches can be loaded with any $C$ files, we have the bound: 
\begin{eqnarray}\label{int-alloc}
 |\mathcal{Y}| \leq \binom{N}{C}^{m} \leq \bigg(\frac{Ne}{C}\bigg)^{mC},
 \end{eqnarray}
where the last inequality is a standard upper bound for binomial coefficients. Finally, using \citet{massart2007concentration}'s lemma for Gaussian variables, we have 
\begin{eqnarray} \label{bound1_FTPL}
	\mathcal{G}(\mathcal{Z})\equiv\mathbb{E}_{\bm \gamma} \big[ \max_{\bm z \in \mathcal{Z}} \langle \bm{z}, \bm{\gamma} \rangle \big] = \mathbb{E}_{\bm \gamma} \big[ \max_{\bm z \in \mathcal{Z}_*} \sum_{(i,f): \gamma^i_f >0} z^i_f\gamma^i_f \big] &\leq& \sqrt{dmC}
	\sqrt{2\log |\mathcal Z_*|} \nonumber \\
	&\leq& mC\sqrt{2d\bigg(\log \frac{N}{C} +1\bigg)}.
\end{eqnarray}

\paragraph{Bounding term (c) in Eqn.\ \eqref{FTPL_regret_final}:} Let us denote the file requested by the $i$\textsuperscript{th} user at time $t$ by $f_i.$
Using \citet[Lemma 1.5]{abernethy2016perturbation}, we have
 \begin{eqnarray}\label{hessian_val}
 \big(\grad^2 \Phi_{\eta_t}(\bm{\tilde {X}_t})\big)_{\bm{pq}}= \frac{1}{\eta_t}\mathbb{E}\big[\bm{\hat{z}}_{\bm{p}}\gamma_{\bm{q}}\big],
 \end{eqnarray}
where $\bm{\hat{z}}\in \arg\max_{\bm{z} \in \mathcal{Z}}\langle \bm{z}, \bm{\tilde{X}}_t+\eta_t \bm{\gamma}\rangle,$ and each of the indices $\bm{p,q}$ is a (user, file) tuple. 
 Hence, using Eqn.\ \eqref{hessian_val}, and noting that each user requests only one file at a time, we have:  
\begin{eqnarray}
\bm{x}_t^T \grad^2 \Phi_{\eta_t}(\bm{\tilde {X}_t}) \bm{x}_t &=& \frac{1}{\eta_t}\sum_{i,j \in \mathcal{I}} \mathbb{E}[\hat{z}^i_{f_i} \gamma^j_{f_j} ]\nonumber \\
 &=& \frac{1}{\eta_t}\mathbb{E}(\sum_{i \in \mathcal{I}} \hat{z}^i_{f_i})(\sum_{j \in \mathcal{I}} \gamma^j_{f_j})	\nonumber \\
 & \stackrel{(a)}{\leq} & \frac{1}{\eta_t}\sqrt{\mathbb{E}(\sum_{i \in \mathcal{I}} \hat{z}^i_{f_i})^2\mathbb{E}(\sum_{j \in \mathcal{I}} \gamma^j_{f_j})^2} \nonumber \\
 & \stackrel{(b)}{\leq}& \frac{1}{\eta_t}\sqrt{n^2 \times n} \nonumber \\
 &=& \frac{1}{\eta_t}n^{3/2}, \label{quad_bd}
\end{eqnarray}
where the inequality (a) follows from the Cauchy-Schwartz inequality and the inequality (b) follows from the facts that $\bm{z}$ are binary variables and that the components of the random vector $\bm{\gamma}$ are i.i.d.
Finally, substituting the upper bounds from Eqns.\ \eqref{bound1_FTPL} and \eqref{quad_bd} in the regret upper bound in Eqn.\ \eqref{FTPL_regret_final}, we may upper bound the expected regret of the \texttt{LeadCache} policy as: 
\begin{eqnarray*}
\mathbb{E}\big(R^{\texttt{LeadCache}}_{T}\big) &\leq& \eta_{T+1} \mathcal{G}(\mathcal Z) + \frac{n^{3/2}}{2}\sum_{t=1}^T \frac{1}{\eta_t} \\
&\leq &\eta_{T+1}mC\sqrt{2d\bigg(\log \frac{N}{C} +1\bigg)} +  \frac{n^{3/2}}{2}\sum_{t=1}^T \frac{1}{\eta_t}, 	
\end{eqnarray*}
where the bound in the last inequality follows from Eqn.\ \eqref{bound1_FTPL}. Choosing the learning rates $\eta_t= \beta\sqrt{t}$ with an appropriate constant $\beta >0$ yields the following regret upper bound:
 \begin{eqnarray*}
 	\mathbb{E}\big(R^{\texttt{LeadCache}}_T\big) \leq kn^{3/4}d^{1/4}\sqrt{mCT},
 \end{eqnarray*}
for some $k=O(\textsf{poly-log}(N/C)).$ $\hfill \blacksquare$ 

%

\section{Proof of Theorem \ref{approx_thm}} \label{approx_thm_proof}

Denote the objective function of Problem \eqref{master_opt} by $L(\bm {y}) \equiv \sum_{i,f} \theta^i_f \min(1, \sum_{j \in \partial^+(i)} y^j_f),$ where, to simplify the notations, we have not explicitly shown the dependence of the $\bm{\theta}$ coefficients on the time index $t$. Recall the definition of surrogate objective function $\phi(\bm{y})$ given in Eqn.\ \eqref{surr_def12}:
\begin{eqnarray}\label{surr_def1}
\phi(\bm y)=\sum_{i,f} (\theta^i_f)^+ \big(1- \prod_{j \in \partial^+(i)}(1-y^j_f) \big),
\end{eqnarray}
From \citet[Eqn.\ (16)]{ageev2004pipage}, we have the following algebraic inequality:
\begin{eqnarray} \label{bd_ineq}
L(\bm y)\stackrel{(a)}{\geq} \phi(\bm y) \geq \bigg(1-(1-\frac{1}{\Delta})^\Delta\bigg)L(\bm y),
\end{eqnarray}
where $\Delta \equiv \max_{i \in \mathcal{I}}|\partial^+(i)|.$ Note that inequality (a) holds with equality for binary vectors $\bm{y} \in \{0,1\}^{mN}$.\\ 
Let $\bm{y}^*$ be a solution of the relaxed LP \eqref{mainLP}, and $\textsf{OPT}$ be the optimal value of the problem \eqref{master_opt}. Obviously, $L(\bm{y}^*)\geq \textsf{OPT}$, which, combined with the estimate in Eqn.\ \eqref{bd_ineq}, yields:  
\begin{eqnarray} \label{eq:3}
\phi(\bm{y}^*)  \geq \bigg(1-(1-\frac{1}{\Delta})^\Delta\bigg) \textsf{OPT}. 
\end{eqnarray}
Since $\bm{y}^*$ is a solution to the relaxed LP, it may possibly contain fractional coordinates. In the following, we show that the Pipage rounding procedure, described in Algorithm \ref{pipage}, rounds at least one fractional variable of a cache at a round \emph{without} decreasing the value of the surrogate objective function $\phi(\cdot)$ (Steps \ref{cache_sel}-\ref{rd2}).\\
 For a given fractional allocation vector $\bm{y}$, and another vector $\bm{v}_{\bm{y}}$ of our choice depending on $\bm{y},$ define a one-dimensional function $g_{\bm y}(\cdot)$ as:
 \begin{eqnarray} \label{g_fn}
  g_{\bm{y}}(s)= \phi(\bm y+ s \bm v_{\bm{y}}).
  \end{eqnarray}
The vector $\bm{v}_{\bm{y}}$ denotes the direction along which the fractional allocation vector $\bm{y}$ is rounded in the current step. The Pipage rounding procedure, Algorithm \ref{pipage}, chooses the vector $\bm{v}_{\bm{y}}$ as follows: consider any cache $j$ that has at least two fractional coordinates $y^j_{f_1}$ and $y^j_{f_2}$ in the current allocation $\bm{y}$ (Step \ref{cache_sel} of Algorithm \ref{pipage}) \footnote{Since the cache capacities are integers, there cannot be a cache with only one fractional allocation variable.}. Take $\bm{v}_{\bm{y}}= e_{j,f_1}-e_{j,f_2}$, where $e_{j,l}$ denotes the standard unit vector with $1$ in the coordinate corresponding to the $l$\textsuperscript{th} file of the $j$\textsuperscript{th} cache, $l=f_1, f_2.$ 
We now claim that the function $g_{\bm y}(s) = \phi(\bm{y}+ s \bm{v}_{\bm{y}})$ is linear in $s$. To see this, consider any one of the constituent terms of $g_{\bm{y}}(s)$ as given in Eqn.\ \eqref{surr_def1}. Examining each term, we arrive at the following two cases:
\begin{enumerate}
\item If both $f \neq f_1$ and $f\neq f_2$ then that term is independent of $s$. 
\item If either $f=f_1,$ or $f=f_2$, the variables $y^j_{f_1}$ or $y^j_{f_2}$ may appear in each product term in \eqref{surr_def1} at most once. Since the product terms contain exactly one variable corresponding to each file, the variables $y^j_{f_1}$ and $y^j_{f_2}$ can not appear in the same product term together.  
\end{enumerate}
The above two cases imply that the function $g_{\bm{y}}(s)$ is linear in $s$. By increasing and decreasing the variable $s$ to the maximum extent possible, so that the candidate allocation $\bm{y} + s \bm{v}_{\bm{y}}$ does not violate the constraint \eqref{relaxed_constr1}, we construct two new candidate allocation vectors $\bm{\alpha}= \bm{y} -\epsilon_1 \bm{v}_{\bm{y}} $ and $\bm{\beta}= \bm{y} + \epsilon_2 \bm{v}_{\bm{y}},$ where the constants $\epsilon_1$ and $\epsilon_2$ are chosen in such a way that at least one of the fractional variables of $\bm{y}$ becomes integral (Steps \ref{dir11}-\ref{dir13}). It is easy to see that, by design, all cache capacity constraints in Eqn.\ \eqref{cap_ctr2} continue to hold in both of these two candidate allocations. In step \ref{rd2}, we choose the best of the candidate allocations $\bm{\alpha}$ and $\bm{\beta},$ corresponding to the surrogate function $\phi(\cdot).$ Let $\bm{y}^{\textrm{new}}$ denote the new candidate allocation vector.
Since the maximum of a linear function over an interval is achieved on one of its two boundaries, we conclude that $\phi(\bm{y}^{\textrm{new}}) \geq  \phi(\bm{y})$. As argued above, the rounded solution is feasible and has at least one less fractional coordinate. Hence, by repeated application of the above procedure, we finally arrive at a feasible integral allocation $\hat{\bm{y}}$ such that:
\[L(\bm{\hat{y}}) = \phi (\bm {\hat{y}}) \geq \phi(\bm{y}^*)\geq \bigg(1-(1-\frac{1}{\Delta})^\Delta\bigg)\textsf{OPT}, \]
where the first equality follows from that fact that the functions $\phi(\bm y)=L(\bm y)$ on integral points. $\hfill \blacksquare$\\
%


\section{Madow's Sampling Scheme} \label{madow-pseudocode}
Madow's sampling scheme is a simple statistical procedure for randomly sampling a subset of items without replacement from a larger universe with a specified set of inclusion probabilities \citep{madow1949theory}. The pseudocode for Madow's sampling scheme is given in Algorithm \ref{madow}. It samples $C$ items without replacement from a universe with $N$ items such that the item $i$ is included in the sampled set with probability $p_i, 1\leq i \leq N.$ The inclusion probabilities satisfy the feasibility constraint given by Eqn.\ \eqref{feasibility-constraint}.
\begin{algorithm}[H] 
\caption{Madow's Systematic Sampling Scheme without Replacement}
\label{madow}
\begin{algorithmic}[1]
\REQUIRE A universe $[N]$ of size $N$, cardinality of the sampled set $C$, marginal inclusion probability vector $\bm{p} = (p_1, p_2, \ldots, p_N)$ satisfying the feasibility condition \eqref{feasibility-constraint},  
\ENSURE A random set $S$ with $|S|=C$ such that, $\mathbb{P}(i \in S)=p_{i}, \forall i\in [N]$ 

\STATE Define $\Pi_0=0$, and $\Pi_i= \Pi_{i-1}+p_{i}, \forall 1\leq i \leq N.$
\STATE Sample a uniformly distributed random variable $U$ from the interval $[0,1].$
\STATE $S \gets \phi$
\FOR {$i\gets 0$ to $C-1$}
\STATE Select element $j$ if $\Pi_{j-1} \leq U + i < \Pi_j.$ 
\STATE $S \gets S \cup \{j\}.$
\ENDFOR
\STATE \textbf{Return} $S$
\end{algorithmic}
\end{algorithm}
\paragraph{Correctness:}
The correctness of Madow's sampling scheme is easy to verify. Due to the feasibility condition \eqref{feasibility-constraint}, Algorithm \ref{madow} selects exactly $C$ elements. Furthermore,  the element $j$ is selected if the random variable $U$ falls in the interval $ \sqcup_{i=1}^{N}[\Pi_{j-1} -i, \Pi_j-i ).$ Since $U$ is uniformly distributed in $[0,1]$, the probability that the element $j$ is selected  is equal to $\Pi_j - \Pi_{j-1} = p_j, \forall j\in [N].$

\section{Proof of Proposition \ref{relaxed-regret-bound}} \label{relaxed-regret-bound-proof}
The proof of the regret bound with the relaxed action set $\mathcal{Z}_{\textrm{rel}}$ follows the same line of arguments as the proof of Theorem \ref{FTPL_inelastic_th} with integral cache allocations. In particular, we decompose the regret bound as in Eqn.\ \eqref{FTPL_regret_final}, with the difference that we now replace the feasible set $\mathcal{Z}$ in term (b) with the relaxed feasible set $\mathcal{Z}_{\textrm{rel}}$. Observe that, for bounding the term (c), we did not exploit the fact that the variables are integral. Hence, the bound \eqref{quad_bd} holds in the case of the relaxed feasible set as well. However, for bounding the Gaussian complexity in term (b) in the proof of Theorem \ref{FTPL_inelastic_th}, we explicitly used the fact that the cache allocations (and hence, the virtual actions) are integral (\emph{viz.} the counting argument in Eqn. \eqref{int-alloc}). To get around this issue, we now give a different argument for bounding the Gaussian complexity in term (b) for the relaxed action set $\mathcal{Z}_{\textrm{rel}}.$ Note that for any feasible $(\bm{z}, \bm{y}) \in \mathcal{Z}_{\textrm{rel}},$ we have 
\begin{eqnarray} \label{l1-norm-bd}
 ||\bm{z}||_1 = \sum_{i,f} z^i_f \leq \sum_i \sum_{j \in \partial^+(i),f} y_{f}^j \leq d \sum_{j} \sum_f y^j_f \leq mCd,
\end{eqnarray}  
where we have used the fact that each cache is connected to at most $d$ users and that each cache can hold $C$ files at a time. Hence, 
we have
\begin{eqnarray} \label{gauss-comp-bd}
\mathcal{G}(\mathcal{Z}_{\textrm{rel}}) = \mathbb{E}_{\bm{\gamma}}\big[ \max_{\bm{z} \in \mathcal{Z}_{\textrm{rel}}} \langle \bm{z}, \bm{\gamma} \rangle\big] \stackrel{\textrm{(H\"older's ineq.)}}{\leq} \mathbb{E}_{\bm{\gamma}}\big[ \max_{\bm{z} \in \mathcal{Z}_{\textrm{rel}}} || \bm{z}||_1 ||\bm{\gamma} ||_{\infty} \big] 	\leq mCd \sqrt{4 \ln(Nn)},
\end{eqnarray}
where, in the last inequality, we have used the $\ell_1$-norm bound \eqref{l1-norm-bd} 
along with a standard upper bound on the expectation of the maximum of the absolute value of a set of i.i.d. standard Gaussian random variables \citep{wainwright2019high}.    
Now proceeding similarly as in the proof of the regret bound for the action set $\mathcal{Z}$, we conclude that with an appropriate learning rate sequence, we have the following regret upper bound for the relaxed action set $\mathcal{Z}_{\textrm{rel}}:$
\begin{eqnarray*}
\mathbb{E}(\tilde{R}_T^{\textrm{LeadCache}}) \leq \kappa_1 n^{3/4} \sqrt{dmCT},	
\end{eqnarray*}
for some polylogarithmic factor $\kappa_1. ~~~ \blacksquare$

\section{Proof of Theorem \ref{fetch}} \label{fetch_proof}
\paragraph{Discussion:}
To intuitively understand why the total number of fetches is expected to be small under the \texttt{LeadCache} policy, consider the simplest case of a single cache with a single user \citep{SIGMETRICS20}. At every slot, the \texttt{LeadCache} policy populates the cache with a set of $C$ files having the highest perturbed cumulative count of requests $\bm{\Theta}(t)$. For the sake of argument, assume that the learning rate $\eta_t$ is time-invariant. Since at most one file is requested by the user per slot, only one component of $\bm{\Theta}(t)$ changes at a slot, and hence, the \texttt{LeadCache} policy fetches at most \emph{one} new file at any time slot. Surprisingly, the following rigorous argument proves a far stronger result: the total number of fetches over an infinite time interval remains almost surely finite, even with a time-varying learning rate in the network caching setting. 

\textbf{Proof:}
Recall that, under the \texttt{LeadCache} policy, the optimal virtual caching configuration $\bm{z}_t$ for the $t$\textsuperscript{th} slot is obtained by solving the optimization problem $\textsf{P}$:
\begin{eqnarray} \label{lp2}
\max_{\bm{z} \in \mathcal{Z}}  \sum_{i \in \mathcal{I}} \langle \bm{\theta}^i(t), \bm{z}^i \rangle,
\end{eqnarray}
where we assume that the ties (if any) are broken according to some \emph{fixed} tie-breaking rule. As discussed before, the corresponding physical cache configuration $\bm{y}_t$ may be obtained using the mapping $\psi(\cdot)$. 
Now consider a static virtual cache configuration $\bm{\tilde{z}}$ obtained by  replacing the perturbed-count vectors $\bm{\theta}^i(t)$ in the objective function \eqref{lp2} with the vectors $\bm{p}^i, \forall i \in \mathcal{I}$, where $\bm{p}=\big(\bm{p}^i, i\in \mathcal{I}\big)$ is defined to be the vector of long-term file-request probabilities, given by Eqn.\ \eqref{renewal_conc}. In other words, 
\begin{eqnarray}\label{stat_opt}
\tilde{\bm z} \in \arg\max_{\bm{z} \in \mathcal{Z}}\sum_{i \in \mathcal{I}} \langle \bm{p}^i, \bm{z}^i \rangle.
 \end{eqnarray}
 Since the set of all possible virtual caching configurations $\mathcal{Z}$ is finite, the objective value corresponding to any other non-optimal caching configuration must be some non-zero gap $\delta > 0$ away from that of an optimal configuration. Let us denote the set of all \emph{sub-optimal} virtual cache configuration vectors by $\mathcal{B}.$ Hence, for any $\bm{z} \in \mathcal{B},$ we must have:
 \begin{eqnarray}\label{opt_gap}
 	\sum_{i \in \mathcal{I}} \langle \bm{p}^i, \bm{\tilde{z}}^i \rangle \geq \sum_{i \in \mathcal{I}} \langle \bm{p}^i, \bm{z}^i \rangle + \delta.
 \end{eqnarray}

  Let us define an ``error" event $\mathcal{E}(t)$ to be event such  that the \texttt{LeadCache} policy yields a sub-optimal virtual cache configuration (and possibly, a sub-optimal physical cache configuration \eqref{virtual_physical}) at time $t$. Let $G$ be a zero-mean Gaussian random variable with variance $2Nn$. We now upper bound the probability of the error event $\mathcal{E}(t)$ as below:
\begin{eqnarray*}
&&\mathbb{P}(\mathcal{E}(t))\\
&\stackrel{(a)}{\leq} & \mathbb{P} \bigg(\sum_{i \in \mathcal{I}} \langle \bm{\theta}^i(t), \bm{z}^i(t) \rangle > \sum_{i \in \mathcal{I}} \langle \bm{\theta}^i(t), \bm{\tilde{z}}^i \rangle, \bm{z}(t) \in \mathcal{B}	\bigg)\\
&\stackrel{(b)}{\leq} & \mathbb{P} \bigg(\eta_tG \geq  \sum_{i \in \mathcal{I}} \langle \bm{X}^i(t), \bm{\tilde{z}}^i \rangle - \sum_{i \in \mathcal{I}} \langle \bm{X}^i(t), \bm{z}^i(t) \rangle, \bm{z}(t) \in \mathcal{B}	\bigg)\\
&\stackrel{(c)}{=}&  \mathbb{P} \bigg(\eta_t G \geq  \sum_{i \in \mathcal{I}} \langle \bm{X}^i(t), \bm{\tilde{z}}^i \rangle - \sum_{i \in \mathcal{I}} \langle \bm{X}^i(t), \bm{z}^i(t) \rangle, \sum_{i \in \mathcal{I}} \langle \bm{X}^i(t), \bm{\tilde{z}}^i \rangle - \sum_{i \in \mathcal{I}} \langle \bm{X}^i(t), \bm{z}^i(t) \rangle > \frac{\delta t}{2}, \bm{z}(t) \in \mathcal{B}	\bigg)\\
&&+ \mathbb{P} \bigg(\eta_t G \geq  \sum_{i \in \mathcal{I}} \langle \bm{X}^i(t), \bm{\tilde{z}}^i \rangle - \sum_{i \in \mathcal{I}} \langle \bm{X}^i(t), \bm{z}^i(t) \rangle, \sum_{i \in \mathcal{I}} \langle \bm{X}^i(t), \bm{\tilde{z}}^i \rangle - \sum_{i \in \mathcal{I}} \langle \bm{X}^i(t), \bm{z}^i(t) \rangle \leq \frac{\delta t}{2}, \bm{z}(t) \in \mathcal{B}	\bigg)\\
&\stackrel{(d)}{\leq} & \mathbb{P}\bigg(\eta_tG \geq \frac{\delta t}{2}\bigg)+ \mathbb{P}\bigg( \frac{1}{t}\sum_{i \in \mathcal{I}}\langle \bm{X}^i(t), \tilde{\bm z}^i - \bm{z}^i(t) \rangle \leq \frac{\delta}{2}, \bm{z}(t) \in \mathcal{B} \bigg) \\
& = & \mathbb{P}\bigg(\eta_tG \geq \frac{\delta t}{2}\bigg)+ \mathbb{P}\bigg(\sum_{i \in \mathcal{I}}\langle \frac{\bm{X}^i(t)}{t} - \bm{p}^i, \tilde{\bm z}^i - \bm{z}^i(t) \rangle + \sum_{i \in \mathcal{I}} \langle \bm{p}^i, \tilde{\bm z}^i - \bm{z}^i(t) \rangle \leq \frac{\delta}{2}, \bm{z}(t) \in \mathcal{B} \bigg) \\
&\stackrel{(e)}{\leq}& \mathbb{P}\bigg(\eta_tG \geq \frac{\delta t}{2}\bigg)+ \mathbb{P}\bigg(\sum_{i \in \mathcal{I}}\langle \frac{\bm{X}^i(t)}{t} - \bm{p}^i, \tilde{\bm z}^i - \bm{z}^i(t) \rangle \leq - \frac{\delta}{2}\bigg)\\
& \stackrel{(f)}{\leq}& \mathbb{P}\bigg(\eta_tG \geq \frac{\delta t}{2}\bigg) + \mathbb{P}\bigg( \sum_{i \in \mathcal{I}} \sum_{f \in [N]} \bigg|\frac{\bm{X}^i_f(t)}{t}-p^i_f\bigg| \geq \frac{\delta}{2}  \bigg) \\
\end{eqnarray*}
\begin{eqnarray*}
& \stackrel{(g)}{\leq}& \mathbb{P}\bigg(\eta_tG \geq \frac{\delta t}{2}\bigg) + \mathbb{P} \bigg( \bigcup_{i,f} \bigg|\frac{\bm{X}^i_f(t)}{t}-p^i_f\bigg| \geq \frac{\delta}{2Nn}\bigg) \\
& \stackrel{(h)}{\leq} & \mathbb{P}\bigg(\eta_tG \geq \frac{\delta t}{2}\bigg) + \sum_{i,f} \mathbb{P} \bigg(\bigg|\frac{\bm{X}^i_f(t)}{t}-p^i_f\bigg| \geq \frac{\delta}{2Nn}\bigg) \\
& \stackrel{(i)}{\leq}& \exp(-ct)+ Nn \alpha_{\epsilon}(t).
\end{eqnarray*}
  for some positive constants $c$ and $\epsilon,$ which depend on the problem parameters. 
  
  In the above chain of inequalities:
  
(a) follows from the fact that on the error event $\mathcal{E}(t),$ the virtual cache configuration vector $\bm z(t)$ must be in the sub-optimal set $\mathcal{B}$ and, by definition, it must yield more objective value in \eqref{lp2} than the optimal virtual cache configuration vector $\tilde{\bm z},$ \\
(b) follows by writing $\bm{\Theta}(t)= \bm{X}(t)+\eta_t \bm{\gamma},$ and observing that the virtual configurations $\bm{z}(t) \in \mathcal{B}$ and $\bm{\tilde{z}}(t)$ may differ in at most $Nn$ coordinates, and that the normal random variables are increasing (in the convex ordering sense) with their variances, \\
(c) follows from the law of total probability,\\
(d) follows from the monotonicity of the probability measures, \\
(e) follows from Eqn.\ \eqref{opt_gap},\\ 
(f) follows from the fact that for any two equal-length vectors $\bm{a}, \bm{b}, $ triangle inequality yields:
\begin{eqnarray*}
\langle \bm{a}, \bm{b} \rangle \geq -\sum_k |\bm{a}_k| |\bm{b}_k|,	
\end{eqnarray*}
and that $|\bm{\tilde{z}}^i_f - \bm{z}^i_f(t)|\leq 1, \forall i,f,$\\
(g) follows from the simple observation that at least one number in a set of some numbers must be at least as large as the average, \\
(h) follows from the union bound, \\
and finally, the inequality (i) follows from the concentration inequality for Gaussian variables and the concentration inequality for the request process $\{\bm{X}(t)\}_{t\geq 1}$, as given by Eqn.\ \eqref{renewal_conc}.
Using the above bound and the assumptions on the request sequence, we have 
\begin{eqnarray*}
\sum_{t \geq 1} \mathbb{P}(\mathcal{E}(t)) \leq \sum_{t \geq 1} \exp(-ct) + Nn\sum_{t\geq 1} \alpha_\epsilon(t)< \infty.
\end{eqnarray*}
Hence, the first Borel-Cantelli Lemma \cite{} implies that 
\begin{eqnarray*}
\mathbb{P}(\mathcal{E}(t)~~ \textrm{i.o})=0. 	
\end{eqnarray*}
Hence, almost surely, the error events stop after a finite time. Thus, with a fixed tie-breaking rule, the new file fetches stop after a finite time w.p. $1$. $\hfill \blacksquare$

\section{Renewal Processes satisfies the Regularity Condition (A)} \label{renewal_proc}
Suppose that, for any $i \in \mathcal{I}, f \in [N],$ the cumulative request process $\{\bm{X}^i_f(t)\}_{t \geq 1}$ constitutes a renewal process such that the renewal intervals have a common expectation $1/p^i_f$ and a finite fourth moment. Let $S_k$ be the time of the $k$\textsuperscript{th} renewal, $k\geq 1$ \citep{ross1996stochastic}. In other words, the $i$\textsuperscript{th} user requests file $f$ for the $k$\textsuperscript{th} time at time $S_k, k\geq 1.$ Then we have 
\begin{eqnarray*}
\mathbb{P}\bigg(\frac{\bm{X}^i_f(t)}{t}-p^i_f \leq -\epsilon \bigg)	&=& \mathbb{P}\bigg(\bm{X}^i_f(t) \leq t(p^i_f-\epsilon)\bigg) \\
&\leq & \mathbb{P}\bigg(S_{\lfloor t(p^i_f-\epsilon)\rfloor}\geq t\bigg)\\
& \leq & \mathbb{P}\bigg((S_{\lfloor t(p^i_f-\epsilon)\rfloor} - \lfloor t(p^i_f-\epsilon)\rfloor)^4 \geq (t(1-p^i_f+\epsilon))^4\bigg)\\
&\stackrel{(a)}{\leq}& O\big(\frac{1}{t^2}\big),
\end{eqnarray*}
where, in (a), we have used the Markov Inequality along with a standard upper bound on the fourth moment of a centered random walk. Using a similar line of arguments, we can show that 
\begin{eqnarray*}
\mathbb{P}\bigg(\frac{\bm{X}^i_f(t)}{t}-p^i_f \geq \epsilon \bigg)	\leq O(\frac{1}{t^2}). 
\end{eqnarray*}
Combining the above two bounds, we conclude that 
\begin{eqnarray*}
	\sum_{t \geq 1}\mathbb{P}\bigg(|\frac{\bm{X}^i_f(t)}{t}-p^i_f| \geq \epsilon \bigg) < \infty.	
	\end{eqnarray*}
The above derivation verifies the regularity condition A for renewal request processes.$\hfill \blacksquare$

\section{Proof of Theorem \ref{lb_thm}} \label{lower_bound_proof}

We establish a slightly stronger result by proving the announced lower bound for a  regular bipartite network with uniform left-degree $d_L$ and uniform right-degree $d.$ Counting the total number of edges in two different ways, we have $nd_L=md.$ Hence, $d_L= \frac{md}{n}.$ For pedagogical reasons, we divide the entire proof into several logically connected parts.
\paragraph{(A) Some Observations and Preliminary Lemmas:}

To facilitate the analysis, we introduce the following surrogate linear reward function:
\begin{eqnarray}\label{lin_surr}
q_{\textrm{linear}}(\bm{x}, \bm{y})\equiv  \sum_{i \in \mathcal{I}}  {\bm{x}_t^i} \cdot \big(\sum_{j \in \partial^+(i)} \bm{y}_t^j\big). 
\end{eqnarray}
We begin our analysis with the following two observations:
\begin{enumerate}
\item \textbf{Upper Bound:}  From the definition \eqref{reward_definition} of the rewards, we clearly have: 
\begin{eqnarray} \label{obs_1}
  q(\bm{x}, \bm{y}) \leq q_{\textsf{linear}}(\bm{x}, \bm{y}), ~~\forall \bm{x}, \bm{y}. 
 \end{eqnarray}
\item \textbf{Local Exclusivity implies Linearity:} In the case when all caches connected to each user host \emph{different} files, \emph{i.e.,} the cached files are \emph{locally exclusive} in the sense that they are not duplicated from each user's local point-of-view, \emph{i.e.,}
\begin{eqnarray} \label{suff_cond_lin}
	\bm{y}^{j_1} \cdot \bm{y}^{j_2}=0, ~~\forall j_1 \neq j_2: j_1, j_2 \in \partial^+(i), \forall i \in \mathcal{I},
\end{eqnarray}
 the reward function \eqref{reward_definition} reduces to a linear one:
\begin{eqnarray} \label{obs_2}
 q(\bm{x}, \bm{y}) = q_{\textsf{linear}}(\bm{x}, \bm{y}), ~~\forall \bm{x}, \bm{y}. 
 \end{eqnarray}
 \end{enumerate}
The equation \eqref{obs_2} follows from the fact that with the local exclusivity constraint, we have \[\sum_{j \in \partial^+(i)}\bm{y}^j_t \leq \bm{1}, ~~ \forall i\in \mathcal{I},\] where the inequality holds component wise. 
Hence, the `$\min$' operator in the definition of the reward function (Eqn.\ \eqref{reward_definition}) is vacuous in this case. 
To make use of the linearity of the rewards as in Eqn. \eqref{obs_2}, the caches need to store items in such a way that the local exclusivity condition \eqref{suff_cond_lin} holds. Towards this, we now define a special coloring of the nodes in the set $\mathcal{J}$ for a given bipartite graph $\mathcal{G}(\mathcal{I} \cupdot \mathcal{J}, E)$. 

\begin{definition}[Valid $\chi$-coloring of the caches]
Let $\chi$ be a positive integer. A valid $\chi$-coloring of the caches of a bipartite network $\mathcal{G}(\mathcal{I}\cupdot \mathcal{J}, E)$  is an assignment of colors from the set $\{1,2,\ldots, \chi\}$ to the vertices in $\mathcal{J}$ (i.e., the caches) in such a way that all neighboring caches $\partial^{+}(i)$ to every node $i \in \mathcal{I}$ (i.e., the users) are assigned different colors.
\end{definition}

Obviously, for a given bipartite graph $\mathcal{G}$, a valid $\chi$-coloring of the caches exists only if the number of possible colors $\chi$ is large enough. The following lemma gives an upper bound to the value of $\chi$ so that a valid $\chi$-coloring of the caches exists.

\begin{framed}
\begin{lemma} \label{min_N_lemma}
Consider a bipartite network $\mathcal{G}(\mathcal{I} \cupdot \mathcal{J}, E),$ where each user $i \in \mathcal{I}$ is connected to at most $d_L$ caches, and each cache $j \in \mathcal{J}$ is connected to at most $d$ users. Then there exists a valid $\chi$-coloring of the caches where $\chi \leq d_Ld$.
\end{lemma}
 \end{framed}
 \begin{proof}
 From the given bipartite network $\mathcal{G}(V,E)$, construct another graph $H(V',E'),$ where the caches form the  vertices of $H$, \emph{i.e.,} $V' \equiv \mathcal{J}.$ For any two vertices in $u, v \in V',$ there is an edge $(u,v) \in E'$ if and only if a user $i \in \mathcal{I}$ is connected to both the caches $u$ and $v$ in the bipartite network $\mathcal{G}$. Next, consider any cache $j \in \mathcal{J}$. By our assumption, it is connected to at most $d$ users. On the other hand, each of the users is connected to at most $d_L-1$ caches other than $j$. Hence, the degree of any node $j$ in the graph $H$ is upper bounded as: 
 \[ \Delta' \leq d(d_L-1) \leq d_Ld -1. \]
 Finally, using Brook's theorem \cite{diestel2005graph}, we conclude that the vertices of the graph $H$ may be colored using at most $1+\Delta' \leq d_Ld$ different colors. 
   \end{proof}
\paragraph{(B) Probabilistic Method for Regret Lower Bounds:} With the above results at our disposal, we now employ the well-known probabilistic method for proving the regret lower bound \citep{alon2004probabilistic}. The basic principle of the probabilistic method is quite simple. We compute a lower bound to the \emph{expected regret} for any online network caching policy $\pi$ for a chosen joint probability distribution $\bm{p}(\bm{x}_1, \bm{x}_2, \ldots, \bm{x}_T)$ over an \emph{ensemble} of incoming file request sequence. Since the maximum of a set of numbers is at least as large as the expectation, the above quantity also gives a lower bound to the regret for the worst-case file request sequence. Clearly, the tightness of the resulting bound
largely depends on our ability to identify a suitable input distribution $\bm{p}(\cdot)$ that is amenable to analysis and, at the same time, yields a good bound. In the following, we show how this program can be elegantly carried out for the network caching problem.
  
Fix a valid $\chi$-coloring of the caches, and let $k=\chi C$. Consider a library consisting of $N=2k$ different files. We now choose a randomized file request sequence  $\{\bm{x}_t^{i}\equiv \bm{\alpha}_t\}_{t=1}^T$ where each user $i$ requests the same (random) file $\bm{\alpha}_t$ at slot $t$ such that the common file request vector $\bm{\alpha}_t$ is sampled uniformly at random from the set of the \emph{first} $2k$ unit vectors $\{\bm{e}_i \in \mathbb{R}^{2k}, 1 \leq i \leq 2k\}$ independently at each slot \footnote{Recall that, according to our one-hot encoding convention, a request for a file $f$ by any user corresponds to the unit vector $\bm{e}_f.$}. 
Formally, we choose: 
\begin{eqnarray*}
p(\bm{x}_1, \bm{x}_2 , \ldots, \bm{x}_T) := \prod_{t=1}^{T} \bigg(\frac{1}{2k} 1\big(\bm{x}_t^{i_1}=\bm{x}_t^{i_2},~ \forall i_1, i_2 \in \mathcal{I} \big)\bigg).	
\end{eqnarray*}

\paragraph{(C) Upper-bounding the Total Reward accrued by any Online Policy:}
Making use of observation \eqref{obs_1}, the expected total reward $G_T^\pi$ accrued by any online network caching policy $\pi$ may be upper bounded as follows:
\begin{eqnarray}
	G_T^\pi &\leq& \mathbb{E}\bigg( \sum_{t=1}^{T} \sum_{i \in \mathcal{I}} \bm{x}_t^{i} \cdot \sum_{j\in \partial^+(i)}\bm{y}_t^j\bigg) \nonumber \\
	&\stackrel{(a)}{=} & \sum_{t=1}^T \sum_{(i,j) \in E} \mathbb{E} \big(  \bm{x}_t^{i} \cdot \bm{y}_t^j\big) \nonumber \\
	&\stackrel{(b)}{=} & \sum_{t=1}^T \sum_{(i,j) \in E} \mathbb{E} \big(\bm{x}_t^{i} \big) \cdot \mathbb{E}\big(\bm{y}_t^j\big)\nonumber\\
	&\stackrel{(c)}{=}& \frac{1}{2k}\sum_{t=1}^T\mathbb{E} \bigg(\sum_{(i,j) \in E} \sum_{f \in [N]}\bm{y}^j_{tf}\bigg)\nonumber\\
	&\stackrel{(d)}{\leq}& \frac{d}{2k}\sum_{t=1}^T \mathbb{E}\bigg( \sum_{j \in \mathcal{J}} \sum_{f \in [N]} \bm{y}^j_{tf}\bigg)\nonumber\\
	&\stackrel{(e)}{\leq}&  \frac{dmCT}{2k}, \label{inelastic_lb1}
\end{eqnarray}
where the eqn.\ (a) follows from the linearity of expectation, eqn.\ (b) follows from the fact that the cache configuration vector $\bm{y}_t$ is independent of the file request vector $\bm{x}_t$ at the same slot, as the policy is online, the eqn.\ (c) follows from the fact that  each of the $N=2k$ components of the vector $\mathbb{E}\big(\bm{x}_t^i\big)$ is equal to $\frac{1}{2k},$ the inequality (d) follows from the fact that each cache is connected to at most $d$ users, and finally, the inequality (e) follows from the cache capacity constraints.

\begin{figure}[h]
\centering
	\includegraphics[ scale=0.7]{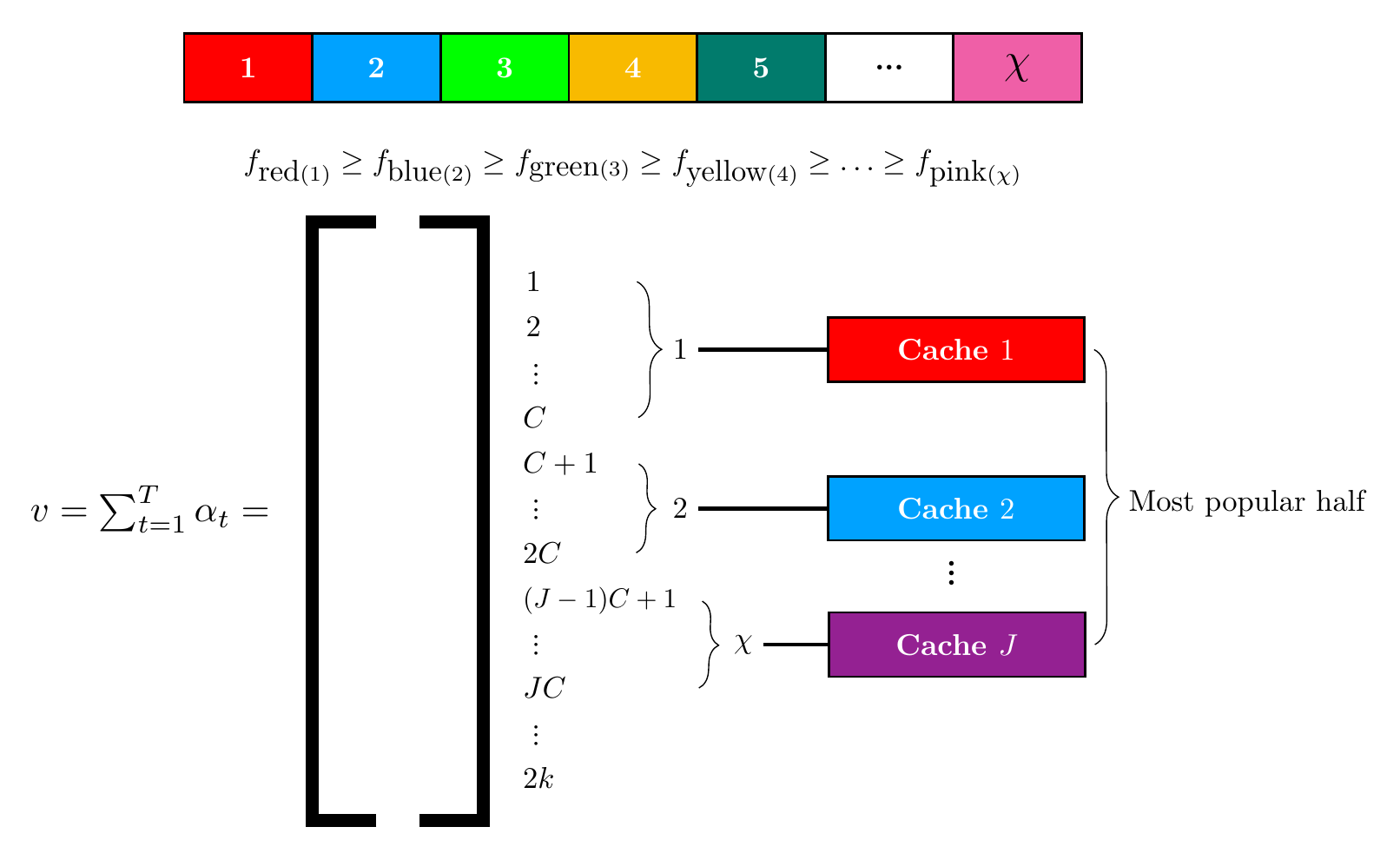}
	\caption{Construction of the caching configuration $\bm{y}_\perp$.
	}
	\label{y_perp}
\end{figure}

\paragraph{(D) Lower-bounding the Total Reward Accrued by the Static Oracle:}
 We now lower bound the total expected reward accrued by the optimal static offline policy (\emph{i.e.,} the first term in the regret definition \eqref{regret_def}). Note that, due to the presence of the `$\min$' operator in \eqref{reward_definition}, obtaining an exactly optimal static cache configuration vector $\bm{y}^*$ is non-trivial, as it requires solving an \textbf{NP-hard} optimization problem \eqref{master_opt} (with the vector $\bm{\theta}(t)$ in the objective replaced by the cumulative file request vector $\bm{X}(T) \equiv \sum_{t=1}^T \bm{x}_t$). However, since we only require a good lower bound to the total expected reward, a suitably constructed sub-optimal caching configuration will serve our purpose, provided that we can compute a lower bound to its expected reward. 
Towards this end, in the following, we construct a joint cache configuration vector $\bm{y}_\perp$ that satisfies the local exclusivity constraint \eqref{suff_cond_lin}.
 \paragraph{D.1 Construction of a ``good" cache configuration vector $\bm{y}_\perp$:}
Let $\mathcal{X}$ be a valid $\chi$-coloring of the caches. Let the color $c$ appear in $f_c$ different caches in the coloring $\mathcal{X}$. To simplify the notations, we relabel the colors in non-increasing order of their frequency of appearance in the coloring $\mathcal{X}$, \emph{i.e.,}  
\begin{eqnarray} \label{color_freq}
f_1 \geq f_2 \geq \ldots \geq f_{\chi}.
\end{eqnarray}
 Let the vector $\bm{v}$ be obtained by sorting the components of the vector $\sum_{t=1}^T \bm{\alpha}_t$ in non-increasing order. Partition the vector $\bm{v}$ into $\frac{2k}{C}=2\chi$ segments by sequentially merging $C$ contiguous coordinates of $\bm{v}$ at a time.  
 Let $c_j$ denote the color of the cache $j \in \mathcal{J}$  in the coloring $\mathcal{X}.$
   The cache configuration vector $\bm{y}_\perp$ is constructed by loading each cache $j \in \mathcal{J}$ with the set of all $C$ files in the $c_j$\textsuperscript{th} segment of the vector $\bm{v}$. See Figure \ref{y_perp} for an illustration.    
   
   \paragraph{D.2 Observation:} Since the vector $\bm{v}$ has $2\chi$ segments (each containing $C$ different files) and the number of possible colors in the coloring $\mathcal{X}$ is $\chi$, it follows that only the most popular half of the files get mapped to some caches under $\bm{y}_\perp$. Moreover, it can be easily verified that the cache configuration vector $\bm{y}_\perp$ satisfies the local exclusivity property \eqref{suff_cond_lin}.
   
   \paragraph{D.3 Analysis:}
 Let $S_{\bm v}(m)$ denote the sum of the frequency counts in the $m$\textsuperscript{th} segment of the vector $\bm{v}$. In other words, $S_{\bm{v}}(m)$ gives the aggregate frequency of requests of the files in the $m$\textsuperscript{th} segment of the vector $\bm{v}.$ By construction, we have 
 \begin{eqnarray}\label{monotonicity}
  S_{\bm{v}}(1) \geq S_{\bm{v}}(2) \geq \ldots \geq S_{\bm{v}}(\chi).
 \end{eqnarray}

Since under the distribution $\bm{p},$ all users  request the same file at each time slot (\emph{i.e.,} $\bm{x}_t^i=\bm{\alpha}_t, \forall i$), and since the linearity in rewards holds due to the local exclusivity property of the cache configuration $\bm{y}_\perp$, the reward obtained by the files in the $j$\textsuperscript{th} cache under the caching configuration $\bm{y}_\perp$ is given by:
\begin{eqnarray}
  \bm{y}_\perp^j\cdot \bigg( \sum_{i \in \partial^-(j)}\sum_{t=1}^{T} \bm{x}_t^i\bigg) &=& \bm{y}_\perp^j\cdot \bigg( \sum_{i \in \partial^-(j)}\sum_{t=1}^{T} \bm{\alpha}_t\bigg) \nonumber \\
  &\stackrel{(a)}{=}& d \bm{y}_\perp^j \cdot \bigg(\sum_{t=1}^{T} \bm{\alpha}_t\bigg) \nonumber \\
  &\stackrel{(b)}=& dS_{\bm v}\big(c_j \big),  \label{inelastic_pf1}
 \end{eqnarray} 
 where the equation (a) follows from the fact that, in this converse proof,  we are investigating a regular bipartite network where each cache is connected to exactly $d$ users, and the equation (b) follows from the construction of the cache configuration vector $\bm{y}_\perp.$
Hence, the expected aggregate reward accrued by the optimal stationary configuration $\bm{y}^*$ may be lower-bounded by that of the configuration $\bm{y}_\perp$ as follows:
\begin{eqnarray} \label{inelastic_star}
 G_T^{\pi^*} &\stackrel{(a)}{\geq}& \mathbb{E}\bigg( \sum_{j \in \mathcal{J}} \bm{y}_\perp^j\cdot \big( \sum_{i \in \partial^-(j)}\sum_{t=1}^{T} \bm{x}_t^i\big) \bigg) \nonumber \\
 &\stackrel{(b)}{=}& d \mathbb{E}\bigg(\sum_{j \in \mathcal{J}}S_{\bm v}\big( c_j\big) \bigg) \nonumber 
 \end{eqnarray}
 \begin{eqnarray*}
  &\stackrel{(c)}{=}& d\mathbb{E}\bigg(\sum_{c=1}^{\chi}f_c S_{\bm{v}}(c) \bigg)\nonumber\\
  &\stackrel{(d)}{\geq} & \frac{d}{\chi} \big(\sum_{c=1}^{\chi} f_c\big)\mathbb{E}\big( \sum_{c=1}^{\chi} S_{\bm v}(c)\big),   
  \end{eqnarray*}
  where \\
  (a) follows from the local exclusivity property of the configuration $\bm{y}_\perp$,\\
  (b) follows from Eqn.\ \eqref{inelastic_pf1}, \\
  (c) follows after noting that the color $c$ appears on $f_c$ different caches in the coloring $\mathcal{X}$, \\
 (d) follows from an algebraic inequality presented in  Lemma \ref{lemma2} below, used in conjunction with the conditions \eqref{color_freq} and \eqref{monotonicity}. 

Next, we lower bound the quantity $\mathbb{E}\big( \sum_{c=1}^{\chi} S_{\bm v}(c)\big)$ appearing on the right hand side of the equation \eqref{inelastic_star}. Conceptually, identify the catalog with $N=2k$ ``bins", and the random file requests $\{\bm{\alpha}_t\}_{t=1}^T$ as ``balls" thrown uniformly into one of the ``bins." With this correspondence in mind, a little thought reveals that the random variable $\sum_{c=1}^\chi S_{\bm{v}}(c)$ is distributed identically as the total load in the most popular $k = \chi C$ bins when $T$ balls are thrown uniformly at random into $2k$ bins. Continuing with the above chain of inequalities, we have
  \begin{eqnarray}\label{final_lb}
  G^{\pi^*}_T&\stackrel{(e)}{\geq}& \frac{dmC}{k}\mathbb{E}\big(\textrm{load in the most popular half of } 2k \textrm{ bins with } T \textrm{ balls thrown u.a.r.}\big) \nonumber\\
  &\stackrel{(f)}{\geq}& \frac{dmC}{k}\bigg( \frac{T}{2}+ \sqrt{\frac{kT}{2\pi}}\bigg)- \Theta(\frac{1}{\sqrt{T}}) \nonumber \\
  &=& \frac{dmCT}{2k} + dmC\sqrt{\frac{T}{2\pi k}} - \Theta(\frac{1}{\sqrt{T}}), \label{inelastic_star2}
\end{eqnarray}
where, in the inequality (e), we have used the fact that $\sum_{c=1}^\chi f_c= m$, and the inequality (f) follows from Lemma \ref{max_load_lemma} stated below. Hence, combining Eqns.\ \eqref{inelastic_lb1} and \eqref{final_lb}, and noting that $k = \chi C \leq d_LdC= \frac{mCd^2}{n}$ from Lemma \ref{min_N_lemma}, we conclude that for any caching policy $\pi:$ 
\begin{eqnarray}\label{lb1}
R^\pi_T \geq G^{\pi^*}_T - G^{\pi}_T 
\geq \sqrt{\frac{mnCT}{2\pi}} - \Theta(\frac{1}{\sqrt{T}}).	
\end{eqnarray} 
Moreover, making use of a \emph{Globally} exclusive configuration (where all caches store different files), in Theorem 7 of their paper, \citet{SIGMETRICS20} proved the following regret lower bound for any online caching policy $\pi$:
\begin{eqnarray}\label{lb2}
	R^\pi_T \geq d \sqrt{\frac{mCT}{2\pi}} - \Theta(\frac{1}{\sqrt{T}}).
\end{eqnarray}
Hence, combining the bounds from Eqns.\ \eqref{lb1} and \eqref{lb2}, we conclude that
\begin{eqnarray*}
R_T^\pi \geq \max \bigg( \sqrt{\frac{mnCT}{2\pi}}, d \sqrt{\frac{mCT}{2\pi}}\bigg) - \Theta(\frac{1}{\sqrt{T}}).	
\end{eqnarray*}
$\hfill \blacksquare$


\begin{framed}
\begin{lemma} \label{lemma2}
Let $f_1\geq f_2 \geq \ldots \geq f_n$ and $s_1 \geq s_2 \geq \ldots \geq s_n$ be two non-increasing sequences of $n$ real numbers each. Then 
\begin{eqnarray*}
\sum_{i=1}^n f_i s_i  \geq \frac{1}{n}\big( \sum_{i=1}^n f_i \big)\big(\sum_{i=1}^n s_i\big).	
\end{eqnarray*}
\end{lemma}
\end{framed}

\begin{proof}
From the rearrangement inequality (\cite{hardy1952inequalities}), we have for each $0\leq j \leq n-1:$ 
\begin{eqnarray}\label{rearrangement}
\sum_{i=1}^n f_is_i \geq \sum_{i=1}^n s_i f_{(i+j)(\textrm{mod }n) + 1},	
\end{eqnarray}
	where the modulo operator is used to cyclically shift the indices. Summing over the inequalities \eqref{rearrangement} for all $0\leq j \leq n-1$, we have 
	\begin{eqnarray*}
	n \sum_{i=1}^n f_is_i \geq \big( \sum_{i=1}^n f_i \big)\big(\sum_{i=1}^n s_i\big),
	\end{eqnarray*}
which yields the result. 
\end{proof}
 \begin{framed}
\begin{lemma}[\cite{SIGMETRICS20}] \label{max_load_lemma}
	Suppose that $T$ balls are thrown independently and uniformly at random into $2C$ bins. Let the random variable $M_C(T)$ denote the number of balls in the most populated $C$ bins. Then
	\begin{eqnarray*}
	\mathbb{E}(M_C(T)) \geq \frac{T}{2} + \sqrt{\frac{CT}{2\pi}} - \Theta\big(\frac{1}{\sqrt{T}}\big).	
	\end{eqnarray*}

\end{lemma}
\end{framed}

\section{Additional Experimental Results} \label{addl_plots}
In this section, we compare the performance of the \texttt{LeadCache} policy (with Pipage rounding) with other standard caching policies on two datasets taken from two different application domains. We observe that the relative performance of the algorithms remains qualitatively the same across the datasets, with the \texttt{LeadCache} policy consistently maintaining the highest hit rate. In our experiments, we instantiated a randomly generated bipartite network with $n=30$ users and $m=10$ caches. Each cache is connected to $d=8$ randomly chosen users. The capacity of each cache is taken to be $10\%$ of the library size. Our experiments are run on HPE Apollo XL170rGen10 Servers with Dual Intel Xeon Gold 6248 20-core and 192 GB RAM.
\subsection{Experiments with the CMU dataset \citep{berger2018practical} }  
\paragraph{Dataset description:} This dataset is obtained from the production trace of a commercial content distribution network. It consists of $500$M requests for a total of $18$M objects. The popularity distribution of the requests follows approximately a Zipf distribution with the parameter $\alpha$ between $0.85$ and $1$. Since we are interested only in the order in which the requests arrive, we ignore the size of the objects in our experiments. Due to the massive volume of the original dataset, we consider only the first $\sim 375$K requests in our experiments.
%
%
%
\begin{table}[!h]
  \caption{Performance Evaluation with the CMU dataset \citep{berger2018practical}}
  \label{sample-table}
  \centering
  \begin{tabular}{llll}
    \toprule
    Policies     &  Hit Rate & Fetch Rate  \\
    \midrule
    \texttt{LeadCache} (with Pipage rounding) & \textbf{0.864}  & \textbf{1.754}      \\
    \texttt{LRU}      & 0.472      &13.375 \\
    \texttt{LFU}    & 0.504    & 13.643 \\
     \texttt{Belady} (offline)    & 0.581     & 5.128  \\
    \bottomrule
  \end{tabular}
  \label{comp-table}
\end{table}

\begin{figure}[h!]
    \centering
    \begin{overpic}[height=2in]{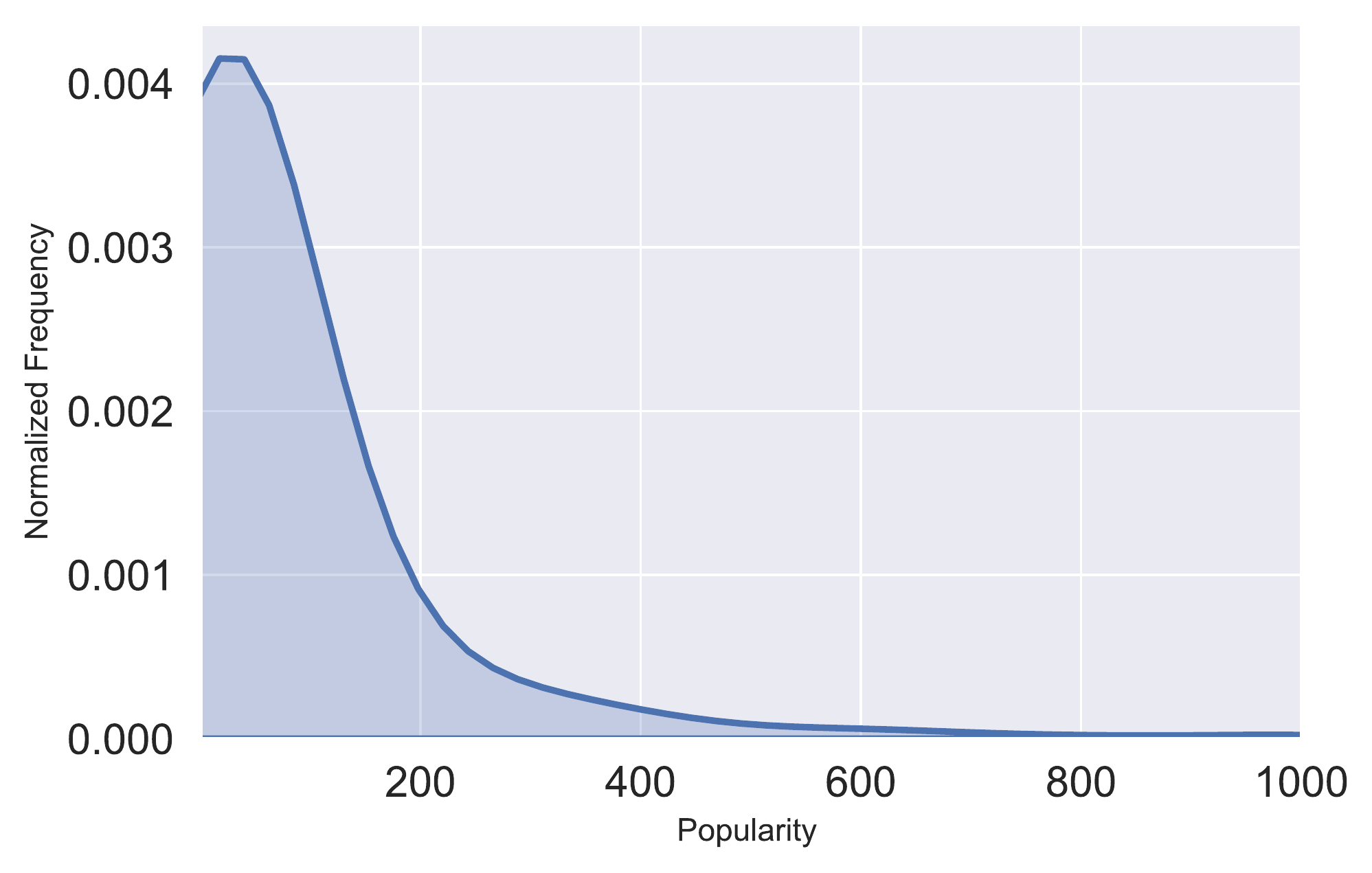}
    \end{overpic}

    \caption{\footnotesize{Plot showing the popularity distribution of the files in the CMU dataset }}
    \label{popularity_plot}
\end{figure}

\begin{figure}
    \centering
    \includegraphics[height=2.1in]{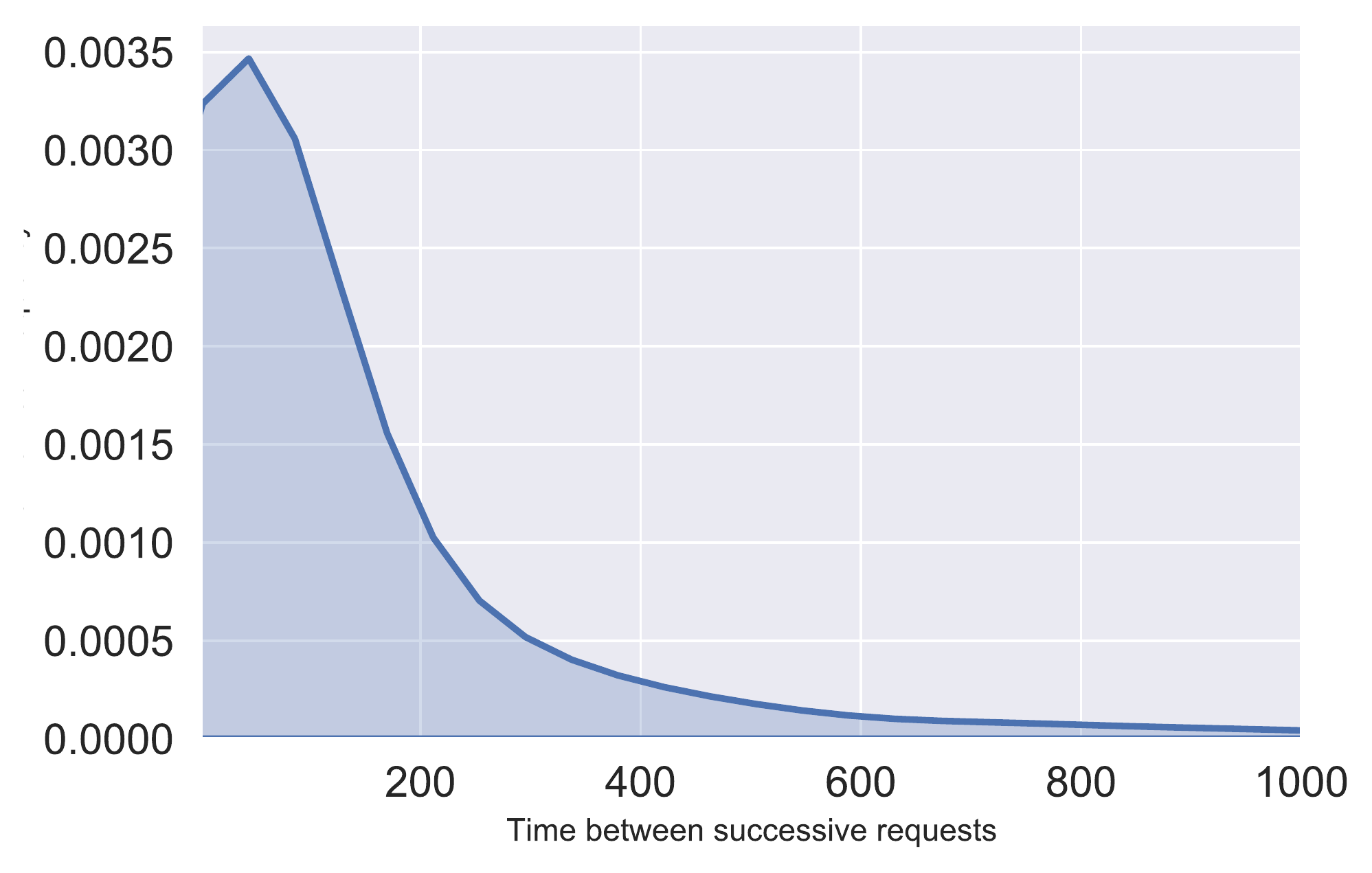}
    \put(-242,40){\rotatebox{90}{\scriptsize{Normalized Frequency}}}
    \caption{Distribution of time between two successive request of the same file on the CMU Dataset}
    \label{inter-req-cmu}
\end{figure}

Figure \ref{popularity_plot} shows the sorted overall popularity distribution of the most popular files in the dataset. It is easy to see that the popularity distribution has a light tail - a small fraction of the files are extremely popular, while others are rarely requested. The \emph{Recall distance} measures the number of file requests between two successive requests of the same file. Figure \ref{inter-req-cmu} shows a plot of the empirical Recall distance distribution for this dataset.  

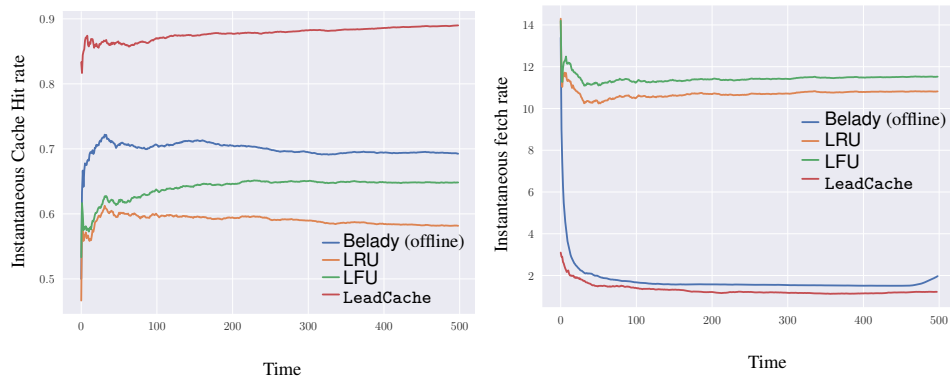
\begin{figure}
\centering
\begin{minipage}{0.45\textwidth}
    \begin{center}
         \begin{overpic}[height=2.1in]{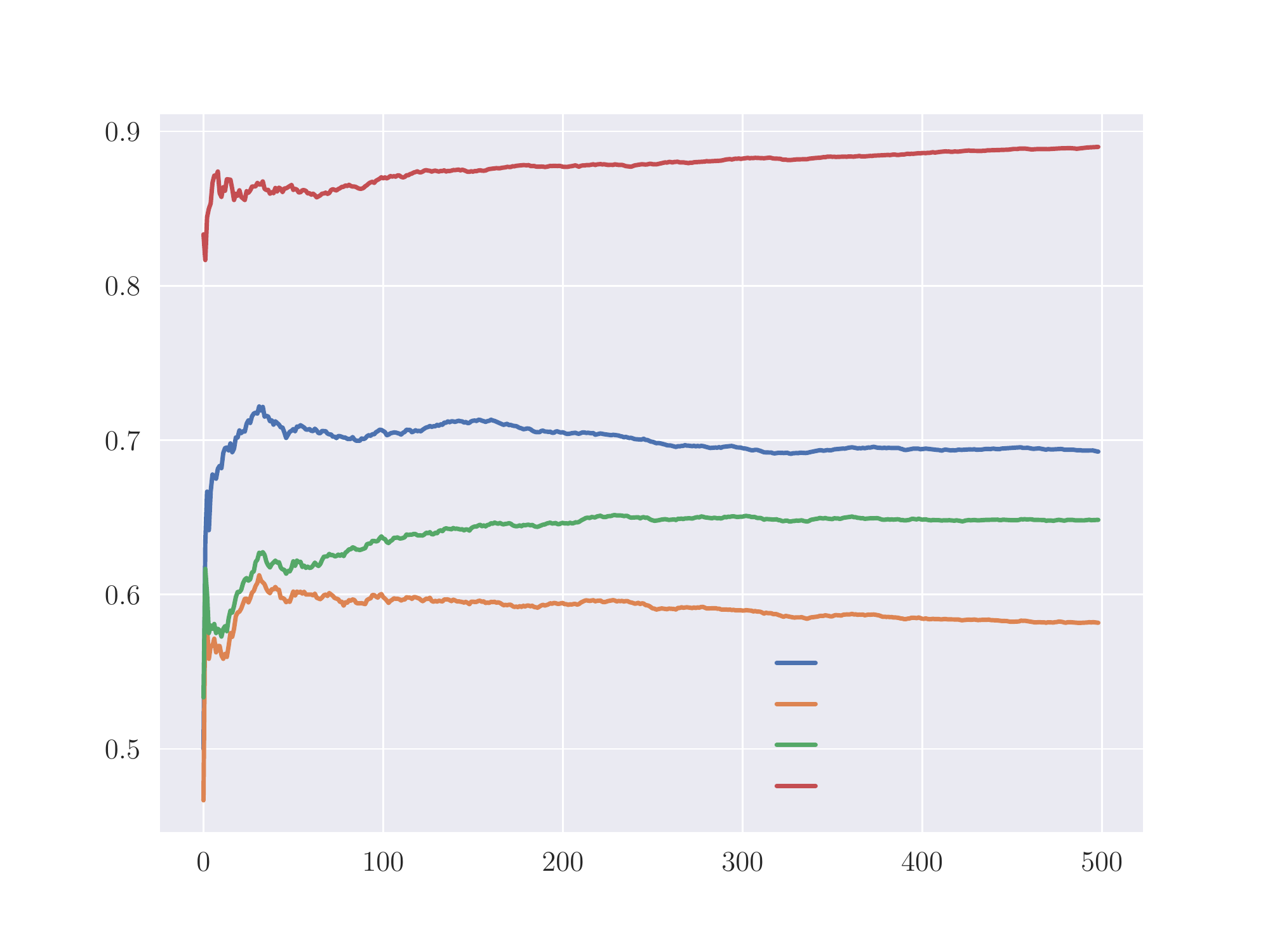}
         \put(65,22){\scriptsize{\textsf{Belady} (offline)}}
         \put(65,18.5){\scriptsize{\textsf{LRU}}}
         \put(65,15){\scriptsize{\textsf{LFU}}}
         \put(65,11.5){\scriptsize{\texttt{LeadCache}}}
         \put(3, 18){\rotatebox{90}{\scriptsize{Instantaneous Cache Hit rate}}}
         \put(50,-2){\scriptsize{Time}}
          \end{overpic}
    \end{center}
   \end{minipage}
  \begin{minipage}{0.45\textwidth}
         \begin{overpic}[height=2.1in]{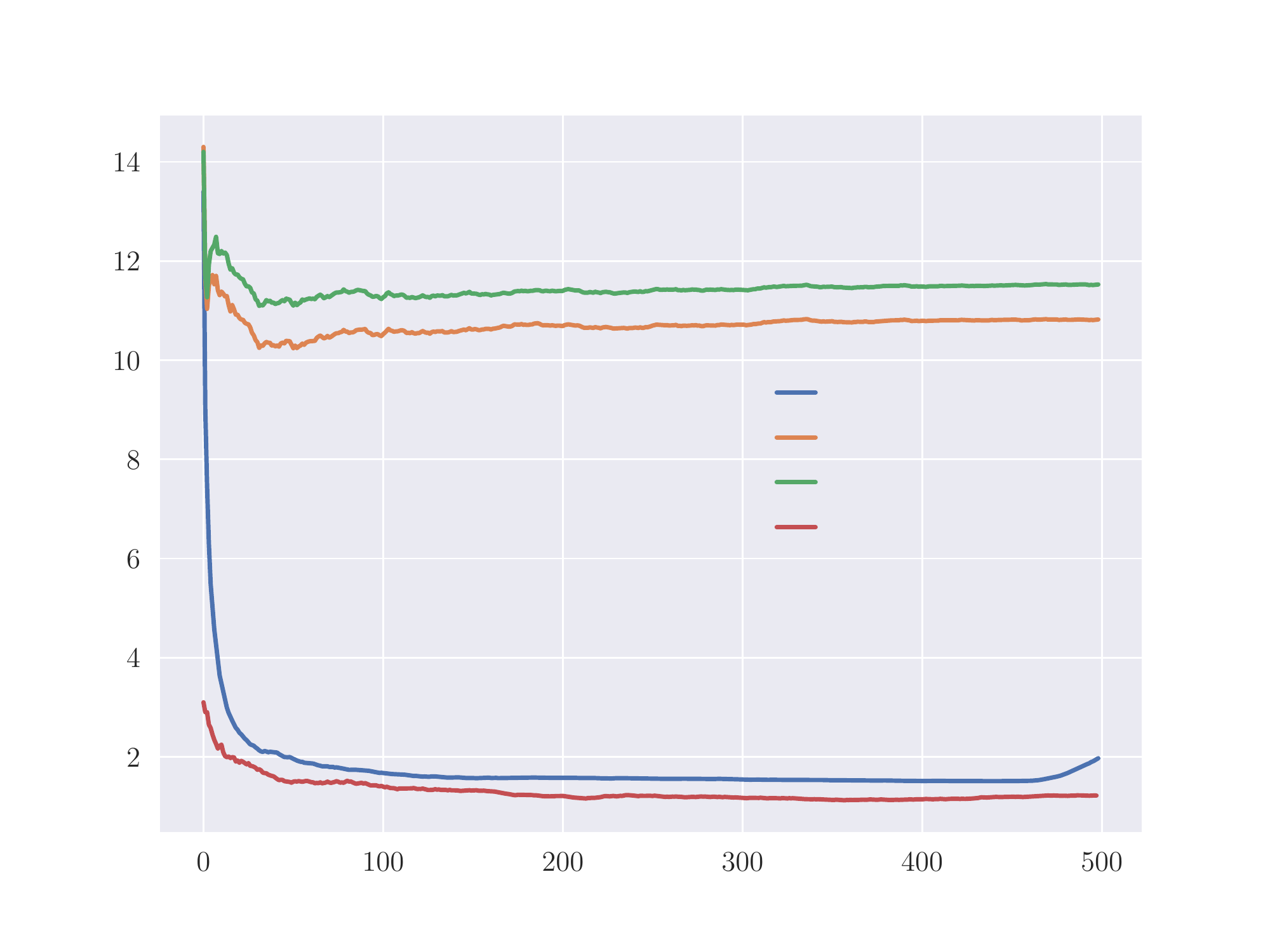}
         \end{overpic}
         \put(-70,88){\scriptsize{\textsf{Belady} (offline)}}
         \put(-70,80){\scriptsize{\textsf{LRU}}}
         \put(-70,72){\scriptsize{\textsf{LFU}}}
         \put(-70,64){\scriptsize{\texttt{LeadCache}}}
         \put(-195, 40){\rotatebox{90}{\scriptsize{Instantaneous fetch rate}}}
         \put(-99,-4){\scriptsize{Time}}

    \end{minipage}
    \caption{\small{Temporal dynamics of instantaneous (a) cache hit rates and (b) fetch rates of different caching policies for a given file request sequence taken from the CMU dataset \citep{berger2018practical}}}
    \label{dyn}
\end{figure}
\begin{figure}
    \centering
    \includegraphics[height=3in]{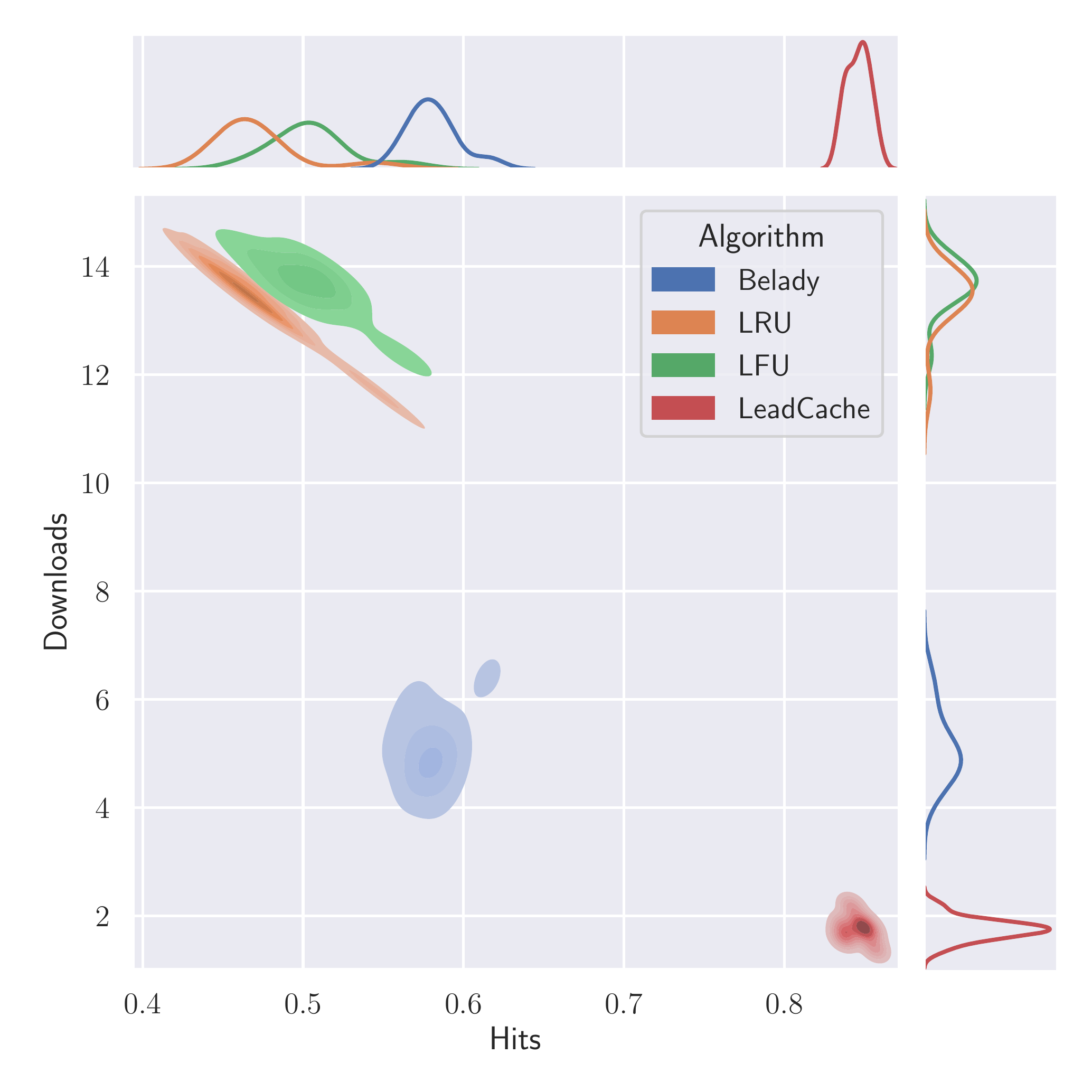}
    \caption{Bivariate plot of cache hit rates and the Fetch rates of different caching policies for the CMU dataset \citep{berger2018practical}}
    \label{joint_plot}
\end{figure}

\paragraph{Experimental Results:}
Figure \ref{dyn} compares the dynamics of the caching policies for a particular file request pattern. It shows that the proposed \texttt{LeadCache} policy maintains a high cache hit rate right from the beginning. In other words, the proposed policy quickly learns the file request pattern from all users and distributes the files near-optimally on different caches. This plot also shows that the fetch rate of the \texttt{LeadCache} policy remains small compared to the other three caching policies. Figure \ref{joint_plot} gives a bivariate plot of the cache hits and downloads by the \texttt{LeadCache} policy. From the plots, it is clear that the \texttt{LeadCache} policy outperforms the benchmarks on this dataset in terms of both Hit rate and Fetch rate. The average values of the performance indices are summarized in Table \ref{comp-table}.

\subsection{Experiments with the MovieLens Dataset \citep{harper2015movielens}}
\begin{figure}
    \centering
    \includegraphics[height=2.1in]{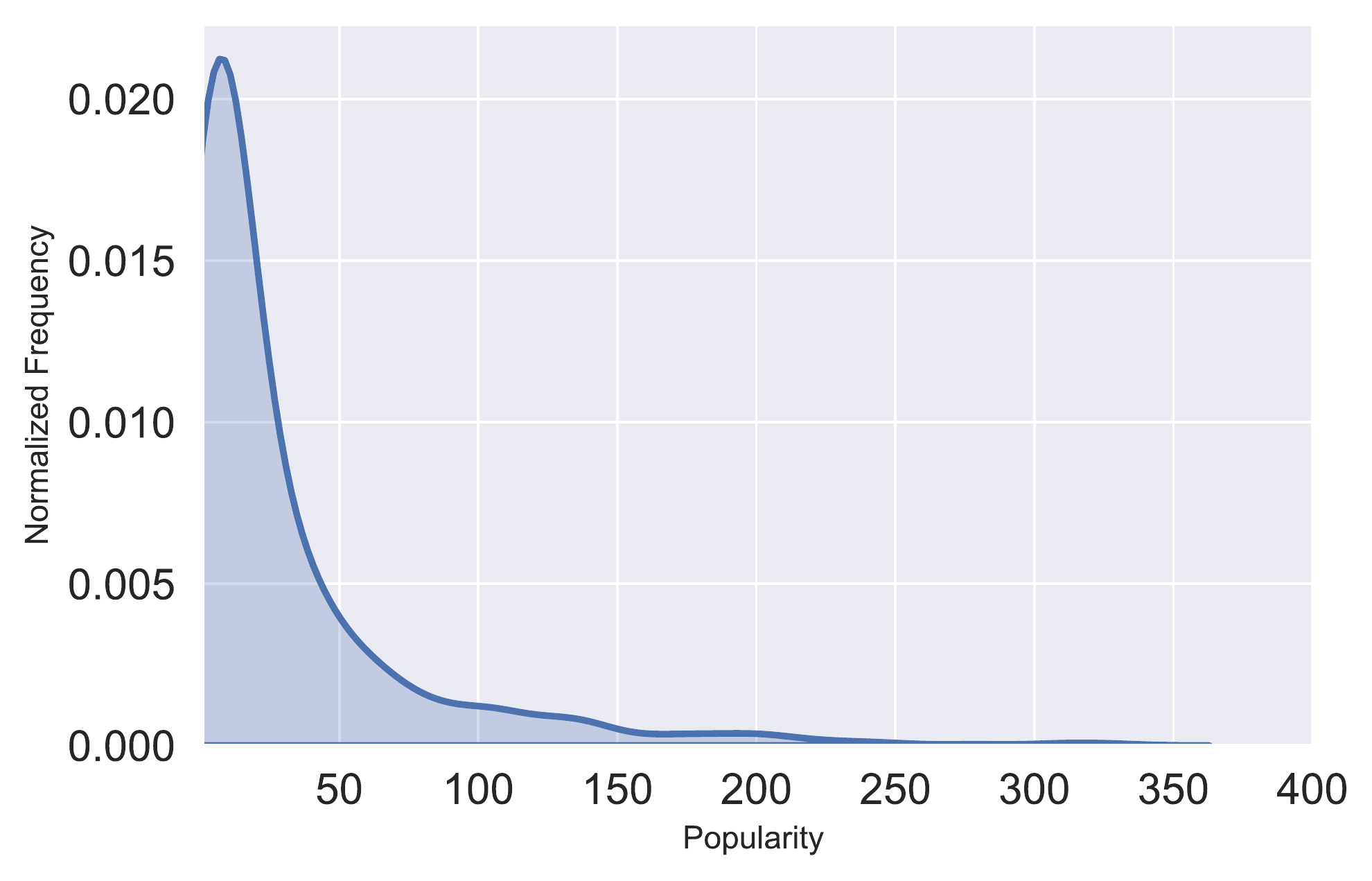}
    \caption{Empirical Popularity distribution of the number of ratings for the MovieLens Dataset \citep{harper2015movielens}}
    \label{popularity-movielens}
\end{figure}

\begin{figure}
    \centering
    \includegraphics[height=2.1in]{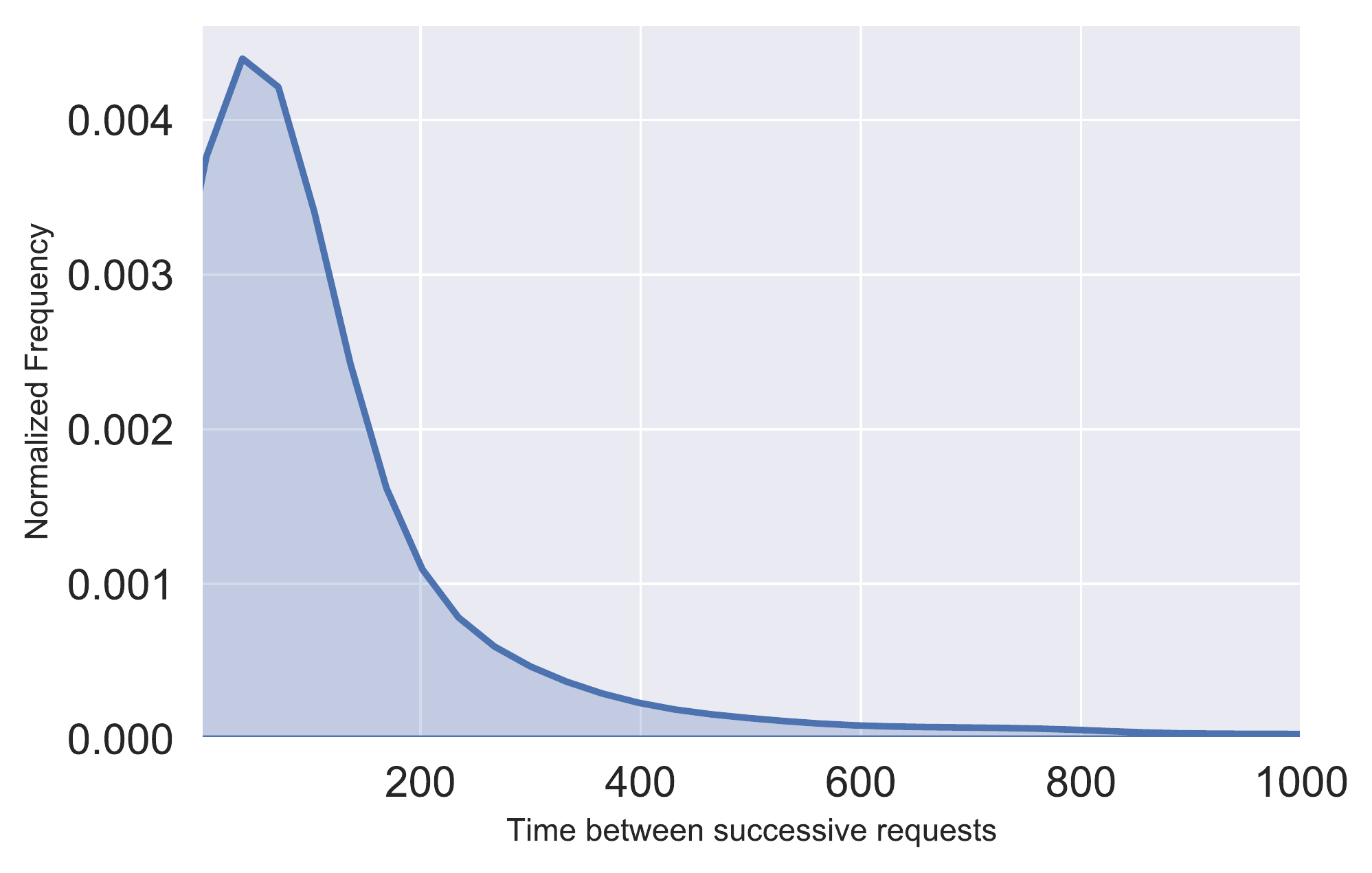}
    \caption{Distribution of time between two successive request of the same file on the MovieLens Dataset}
    \label{inter-req-movielens}
\end{figure}

\paragraph{Dataset Description:} MovieLens \footnote{This dataset is freely available from \url{https://www.kaggle.com/grouplens/movielens-20m-dataset}} is a popular dataset  containing $\sim 20$M ratings for $N \sim 27278$ movies along with the timestamps of the ratings \citep{harper2015movielens}. The ratings were assigned by $138493$ users over a period of approximately twenty years. Our working assumption is that a user rates a movie in the same sequence as she requests the movie file for download from the Content Distribution Network. Due to the sheer size of the dataset, in our experiments, we consider the first 1M ratings only. Figure \ref {popularity-movielens} shows the empirical distribution of the number of times the movies have been rated (and hence, downloaded) by the users. Figure \ref{inter-req-movielens} shows the empirical distribution of time between two successive requests of the same file (\emph{i.e.,} the \emph{Recall distance}). 

\begin{table}[!h]
  \caption{Performance Evaluation with the MovieLens dataset \citep{harper2015movielens}}
  \label{sample-table}
  \centering
  \begin{tabular}{llll}
    \toprule
    Policies     &  Hit Rate & Fetch Rate  \\
    \midrule
    \texttt{LeadCache} (with Pipage rounding) & \textbf{0.991}  & 1.509      \\
    Heuristic \citep{SIGMETRICS20}     & 0.694 & \textbf{0.297}      \\
    \texttt{LRU}      & 0.312      &3.234 \\
    \texttt{LFU}    & 0.595    & 2.028 \\
     \texttt{Belady} (offline)    & 0.560     & 1.589 \\
    \bottomrule
  \end{tabular}
  \label{comp-table-movielens}
\end{table}

\begin{figure}
\centering
\begin{minipage}{0.43\textwidth}
   \hspace{-20pt}
         \begin{overpic}[height=2in]{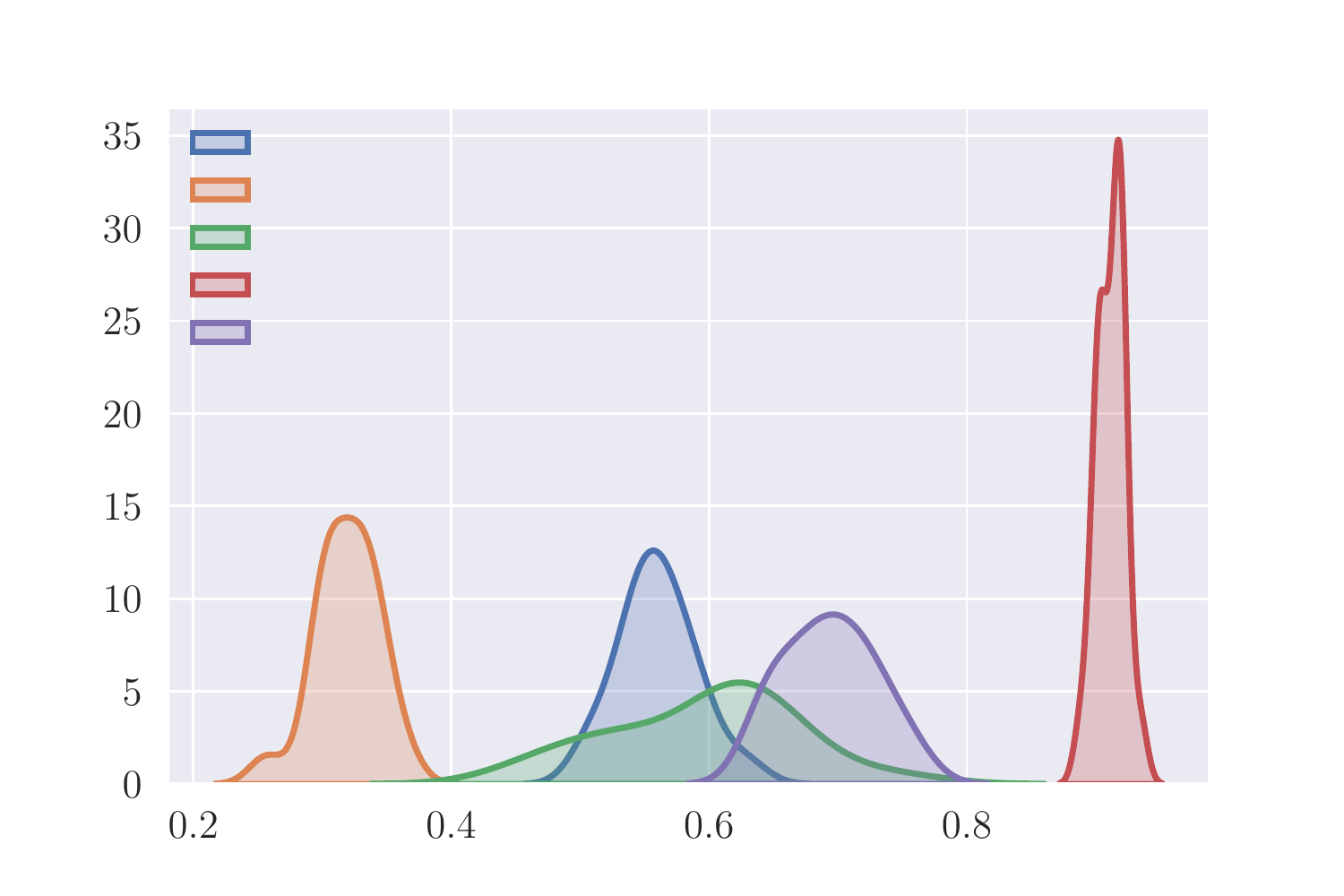}
         \put(20,55){\scriptsize{\textsf{Belady} (offline)}}
         \put(20,51.5){\scriptsize{\textsf{LRU}}}
         \put(20,48){\scriptsize{\textsf{LFU}}}
         \put(20,44.5){\scriptsize{\texttt{LeadCache}}}
          \put(20,41){\scriptsize{\textsf{\citet{SIGMETRICS20} }}}

         \put(1, 18){\rotatebox{90}{\scriptsize{Normalized frequency}}}
         \put(39,-2){\scriptsize{Average cache hit rate}}
          \end{overpic}
 \end{minipage}
 \begin{minipage}{0.43\textwidth}
         \begin{overpic}[height=2in]{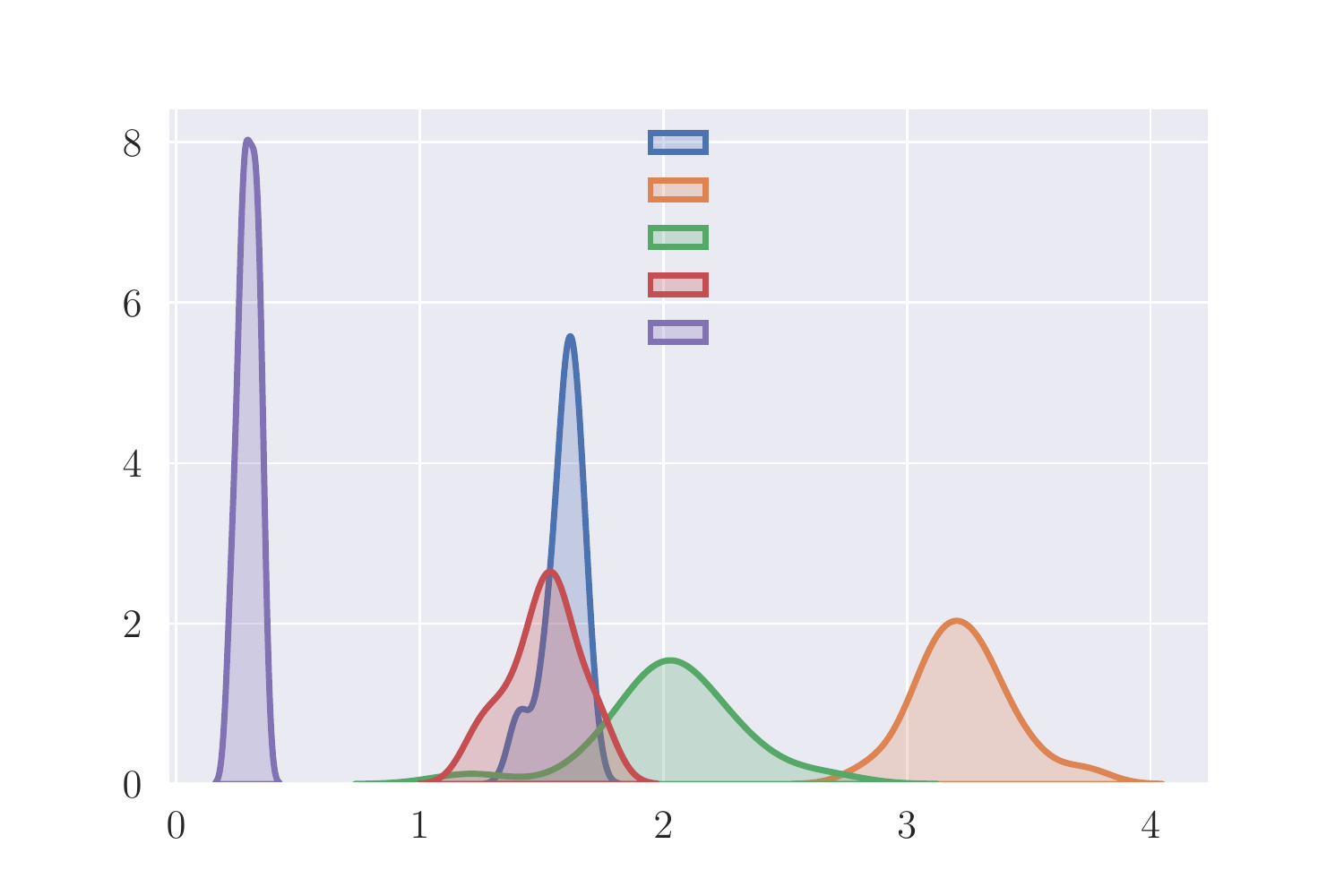}
         \end{overpic}
         \put(-100,120){\scriptsize{\textsf{Belady} (offline)}}
         \put(-100,112){\scriptsize{\textsf{LRU}}}
         \put(-100,104){\scriptsize{\textsf{LFU}}}
         \put(-100,96){\scriptsize{\texttt{LeadCache}}}
         \put(-100,88){\scriptsize{\textsf{\citet{SIGMETRICS20}}}}
         \put(-208, 40){\rotatebox{90}{\scriptsize{Normalized frequency}}}
         \put(-145,-8){\scriptsize{Average download rate per cache}}
    \end{minipage}
    \caption{\small{Empirical distributions of (a) Cache hit rates and (b) Fetch rates of different caching policies on the MovieLens Dataset.}}
    \label{hit-download-rates-movielens}
\end{figure}

\begin{figure}
    \centering
    \includegraphics[height=3in]{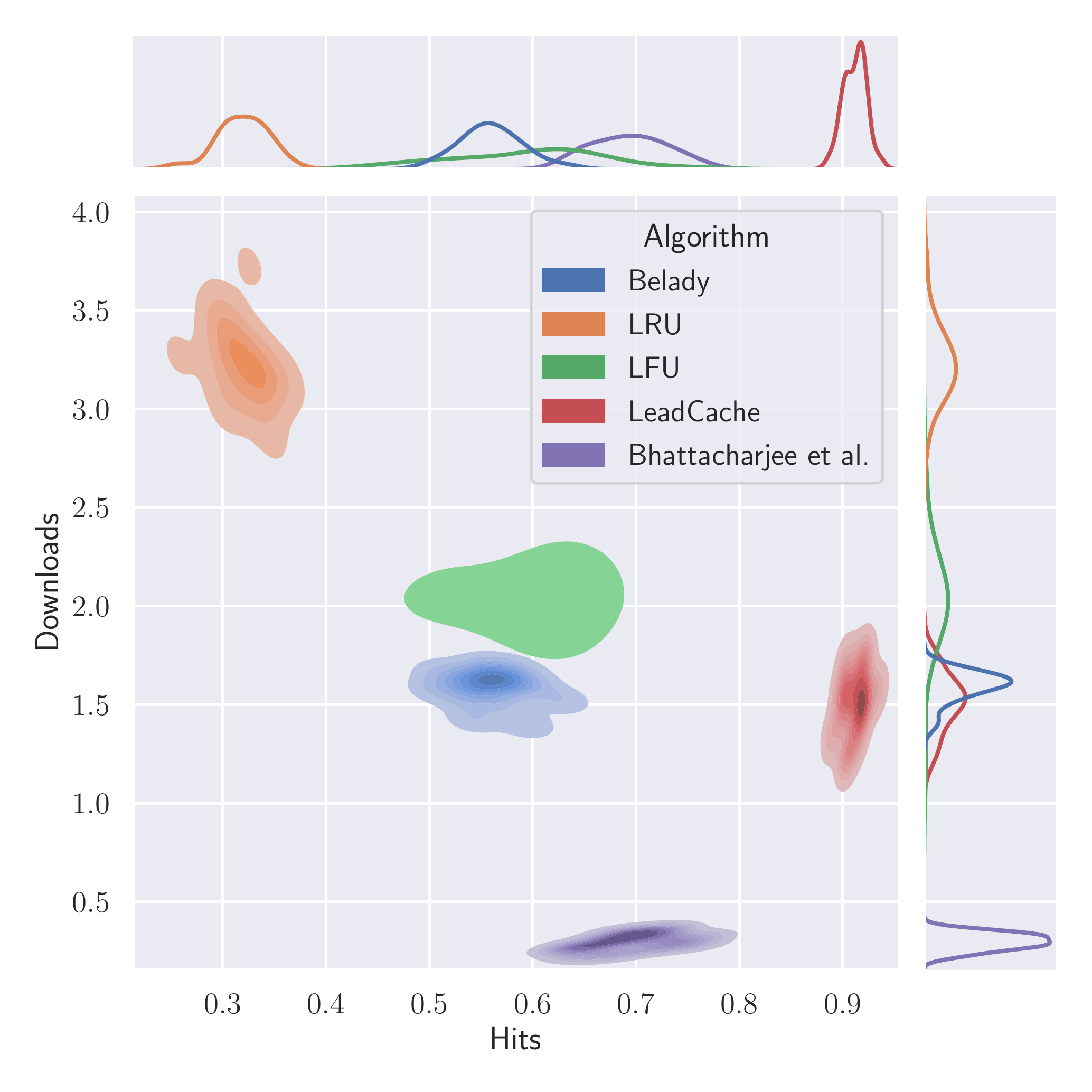}
    \caption{Bivariate plot of cache hit rates and the fetch rates of different caching policies for the MovieLens dataset \citep{harper2015movielens}}
    \label{joint_plot-movielens}
\end{figure} 
\paragraph{Experimental Results} Figure \ref{hit-download-rates-movielens} compares the performance of different policies in terms of the hit rates and fetch rates. The average values of the key performance indicators are shown in Table \ref{comp-table-movielens}. From the plots and the table, we see that the \texttt{LeadCache} policy achieves the highest hit rate among all other policies, which is about $32\%$ more than that of the Heuristic policy proposed by \citet{SIGMETRICS20}. On the other hand, it incurs more file fetches compared to only the heuristic policy proposed by \citet{SIGMETRICS20}. Figure \ref{joint_plot-movielens} gives a joint plot of the hit rate and the fetch rate of different policies. It is clear from the plots that the \texttt{LeadCache} policy robustly learns the file request patterns and caches them on the caches near-optimally.   
%
%
%
%

\clearpage

\end{document}